\def\iems#1{{\protect\index{#1|bold}}}
\def\lqq{``}
\def\rqq{"}
\def\sgn{{\rm sgn}}
 \def\tld{truncated L\`evy distribution}

\def\nablabs{{\scriptsize{\mbox{\boldmath$\nabla$}}}}
\def\deltab{{\mbox{\boldmath$\delta$}}}

\def\aut#1{#1}
\def\dst{\displaystyle}
\def\ins#1{}
\def\iem#1{{\em #1}}

\def\comment#1{}

\def\cm#1{}
\def\sbf#1{\mbox{\scriptsize{\bf #1}}}
\def\ssbf#1{\mbox{\tiny{\bf #1}}}
 \def\lfrac#1#2{{{{#1}/{#2}}}}

\def\>{\rangle}
\def\<{\langle}
\def\comment#1{}
\def\ind#1{#1}

\documentstyle[pra,epsf,aps]{revtex}
\begin{document}

\title{Stochastic Calculus for
Assets with Non-Gaussian Price Fluctuations
\thanks{
kleinert@physik.fu-berlin.de,
{}~ http://www.physik.fu-berlin.de/\~{}kleinert }}
\author{
Hagen Kleinert\\
Institute for Theoretical Physics, FU-Berlin,
Arnimallee 14, D-14195, Berlin, Germany
}
\maketitle
\begin{abstract}
From the path integral formalism
for price fluctuations
with non-Gaussian distributions
I derive the appropriate stochastic calculus
replacing It\^o's calculus for stochastic fluctuations.
\end{abstract}

\section{Introduction}

The logarithms of assets prices
in
financial markets
do not fluctuate with
Gaussian
distributions.
This was first noted by Pareto in
the
 19th century \cite{PAR}
and
emphasized
by
Mandelbrot
 in the 1960s \cite{MB1}.
The true distributions
possess  large  tails
which may be approximated by  various other
distributions, most prominently
the
truncated L\'evy distributions \cite{MB1,MES,KOP,BP}, the Meixner distributions
 \cite{Gri,Mei},
or the generalized hyperbolic distributions
and their descendents
\cite{N1,N2,N2a,N3,N4,N5,N6,N7,N8,N9,N9b,N10,N11,N12,N13,N14,N15,N16,N17,N18,N19,N20,N21,N22,N23}.
The
stochastic differential equations
associated
with such distributions
 cannot be treated with
the popular It\^o calculus.
The purpose of this paper
is to develop an appropriate
calculus to replace it.
In Section \ref{@secG} we briefly  recapitulate the Gaussian
approximation
and set the stage for the generalization
in Section \ref{@LD}.

\section{Gaussian Approximation to
Fluctuation Properties of Stock Prices}
\label{@secG}
\begin{figure}[h]
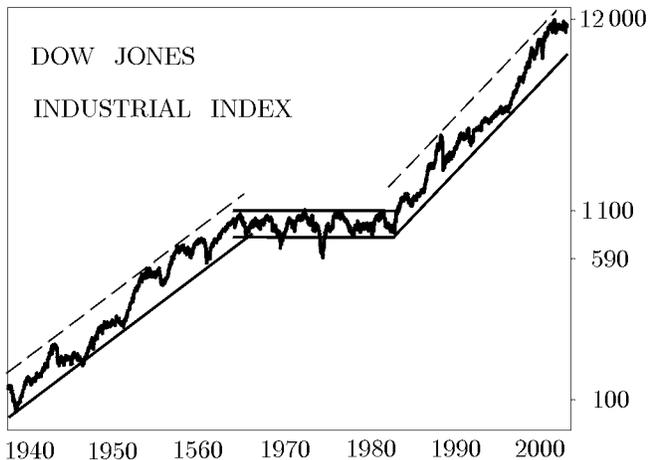

~~~~~~~~~~\input dow.tps
\caption[Periods of exponential growth of
price index averaged
over major industrial stocks
in the Unites States over 60 years]{Periods of exponential growth of
price index averaged over major industrial stocks
in the Unites States over 60 years.}
\label{dowjones}\end{figure}

\begin{figure}[tbhp]
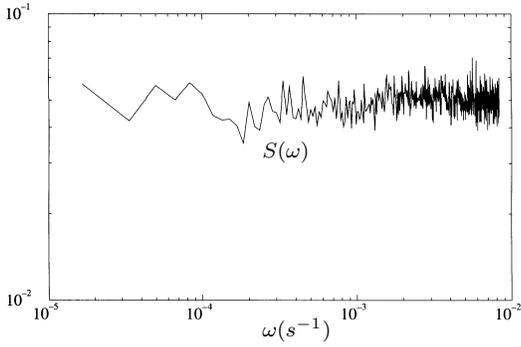

\unitlength .36mm
\input white.tps \\
\caption[Fluctuation
spectrum of the exchange rate
DM/US\$]{Fluctuation
spectrum of exchange rate
DM/US\$ as a function of the frequency, indicating a white noise
in the stochastic differential equation (\ref{@stochdequsto})
(from the textbook \cite{BP}).
}
\label{@white}\end{figure}

Let $S(t)$ denote the price of some stock.
Over long time spans, the average over many
stock prices has a time behavior which can be
approximated by pieces of exponentials.
This is why they are usually
plotted
on  a logarithmic scale.
For an illustration see
 the  Dow-Jones
industrial index
over 60 years in Fig.~\ref{dowjones}.~\\
~\\
For a liquid market with many participants,
the price fluctuations
seem to be driven by a stochastic noise with a
 white spectrum, as illustrated in Fig.~\ref{@white}.

\begin{figure}[hbt]
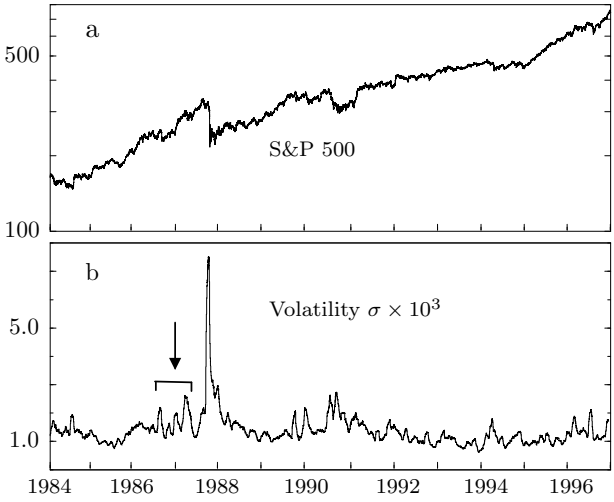

\vspace{-.5cm}
\hspace{-3cm}
\input sundp.tps  \phantom{xxxxxxxxxxxxxxxxxx}
\hspace{1cm}
\vspace{-.6cm}
\caption[(a)  The S\&P 500 index for the 13-year
period 1 Jan 1984 - 31 Dec 1996 at interval of 1 min.
(b)
Volatility with T=1\,mon (8190\,min) and time lag 30\,min.
]
{(a)  The S\&P 500 index for the 13-year
period 1 Jan 1984 - 31 Dec 1996 at interval of 1 min.
Large
fluctuations appeared on 19 Oct 1987 ({\em black Monday\/}). (b)
Volatility over times T=1\,mon (8190\,min)
for courses taken every 30min. The
precursors of the '87 crash are indicated by arrows
(from Ref.~\cite{logno}).}
\label{@s+pvol}\end{figure}

The fluctuations of  the index
have a certain average width called  the \iem{volatility}
of the market.
Over longer time spans, the volatility changes stochastically,
as illustrated by the data of the S\&P500 index over the years
1984-1997 shown in Fig.~\ref{@s+pvol}. In particular, there are strong increases
short before market\linebreak
~\\[-5mm]
crashes.
The theory to be developed here
will ignore
these fluctuations and assume a constant
 volatility.
For recent work taking them into account see
\cite{Baaquie}.

\begin{figure}[tbh]
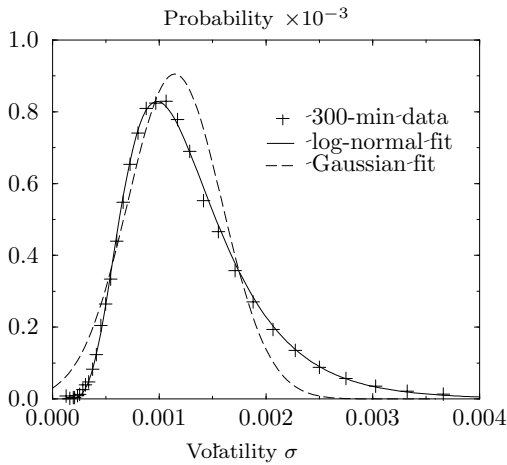
 \vspace{-5mm}
\input lognorvo.tps ~\\
\caption[Comparison of the best log-normal
and Gaussian fits to volatilities
over  300-min
]{Comparison of the best log-normal
and Gaussian fits to volatilities
over  300-min
(from Ref.~\cite{logno}).}
\label{@lognorvo}\end{figure}

The distribution of the logarithms of the volatilities is
normal as
shown in Fig.~\ref{@lognorvo}.

An individual stock will in general have larger  volatility
than an average market index, especially when the company
is small and only few stocks are traded per day.

\begin{figure}[tbhp]
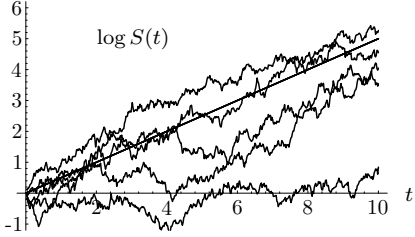

\input stoch0.tps
\caption[Behavior of logarithm of stock price following
the stochastic differential equation
(\protect\ref{@stochdequsto})]{Behavior of logarithm of
stock price following the stochastic differential equation
(\protect\ref{@stochdequsto}).}
\label{@stoch0}\end{figure}

To lowest approximation,
 the
stock price
$
 S(t)
$
 satisfies the simplest stochastic differential equation
for exponential growth
\begin{equation}
\frac{\dot S(t)}{S(t)}=r_S+  \eta (t),
\label{@stochdequsto}\end{equation}
where
$ \eta (t)$ is a white noise
with the correlation functions
\begin{equation}
\langle \eta(t)  \rangle =0,~~~~~~
\langle \eta(t) \eta(t') \rangle =
  \sigma ^2\delta (t-t').
  \label{stocknoise}
\end{equation}
The standard deviation $ \sigma$
parametrizes
the
volatility of the stock price, which is
measured by the expectation value
\begin{eqnarray}
 \Bigg \langle\left[ \frac{\dot S(t)}{S(t)}\right]^2   \Bigg \rangle\,dt
= \sigma ^2            .
\label{@volati}\end{eqnarray}
The
logarithm of the stock price
\begin{equation}
x(t)=\log S(t)
\label{@logf}\end{equation}
does not simply satisfy the stochastic linear-growth
differential equation
$\dot x(t)=\dot S(t)/S(t)=r_S+  \eta (t) $, but possesses another
growth rate $r_x$:
\begin{eqnarray}
\dot x(t)&=&r_{x}+      \eta (t) .
\label{@stochdequstox}\end{eqnarray}
The change
of the
growth rate
is
due the stochastic nature of  $x(t)$ and $S(t)$.
According to  It\^o's rule,
a function of a stochastic variable
satisfies the relation
\begin{eqnarray}
 \dot f( x(t)) =  \partial _xf(x(t))
\dot x(t)
+\frac{ \sigma ^2}{2}f''(x(t)) .
\label{@Itosrule}\end{eqnarray}
This is derived by expanding
$f( x(t+ \epsilon ))$ in a Taylor series
\begin{eqnarray} ~ \!\!\!\!\!\!\!\!\!\!\!
&&~~\!\!\!\!\!\!\!\!\!
\!\!\!\!\!\!\!\!\!\!
f( x(t+ \epsilon ))\!=\!f(x(t)) +
\,f'(x(t))
 \int _t^{t+ \epsilon } dt'\,
 \dot x(t')
\nonumber \\&& ~~~~~~~~~~~~~ ~ ~~~~~ \hspace{1pt}
+~\frac{1}{2}f''(x(t))
\int _t^{t+ \epsilon } dt_1\int _t^{t+ \epsilon }
dt_2\,
 \dot x(t_1)
 \dot x(t_2)
\label{@Itonew200} \\&& ~~~~~~~~~~~~~~~~~~~\hspace{-1pt}
 \,
+~\frac{1}{3!}f^{(3)}(x(t))\int _t^{t+ \epsilon }
dt_1\int _t^{t+ \epsilon }
dt_2\int _t^{t+ \epsilon }
t_3\,\dot x(t_1)\dot x(t_2)\dot x(t_3)
+\dots~  ,  \nonumber
\end{eqnarray}
replacing the higher correlation functions on the right-hand side
by their expectation values, and observing that
only the quadratic noise term in $\dot x(t)=r_x+   \eta (t)$
contributes to linear order in $ \epsilon $.
For the logarithmic function
(\ref{@logf}), the expansion yields
\begin{eqnarray}
\dot x(t)&=&
\frac{dx}{dS}\dot S(t) +\frac{1}{2}\frac{d^2x}{dS^2}\dot S^2(t)\, dt+\dots\nonumber \\
&=&\frac{\dot S(t)}{S(t)}-\frac{1}{2}\left[ \frac{\dot S(t)}{S(t)}\right]^2  dt+\dots
,
\label{@Itosrule0}\end{eqnarray}
which becomes, after the replacement of the quadratic term by its expectation
value (\ref{@volati}),
\begin{eqnarray}
\dot x(t)&=&
\frac{\dot S(t)}{S(t)}-\frac{1}{2} \sigma^2+\dots~.
\label{@Itosrule1}\end{eqnarray}
Inserting here
Eqs.~(\ref{@stochdequsto}),
and
(\ref{@stochdequstox}),
we find
that the
 linear
growth rate
of $x(t)$
is related to the exponential growth rate  of $S(t)$
by
\begin{eqnarray}
r_ {x}=r_S- \frac{1}{2}  \sigma ^2.
\label{@itoreln}\end{eqnarray}
In praxis, this relation implies
that if we fit
a straight line
 through a plot of the logarithms
of stock prices, the forward
extrapolation of the average stock price is
given by
\begin{equation}
\langle S(t)\rangle =S(0)\,e^{r_St}=
S(0)\,e^{(r_{x}+ \sigma ^2/2)t}.
\label{@forwS}\end{equation}
A typical set of solutions of the stochastic differential equation
(\ref{@stochdequstox}) is shown
in Fig.~\ref{@stoch0}.

\section{L\'evy Distributions}
\label{@LD}The description
of the fluctuations
of the logarithms of the stock prices
around the linear trend
by a Gaussian distribution
is  only a rough
approximation to the real
stock prices.
As explained before,
these have
volatilities
depending on time.
More severely,
if observed in smaller time intervals,
for instance every day or hour,
they
have distributions
in
which rare events have a
much higher probability than
in Gaussian
distributions, whose
exponential tails are extremely small.
Following Pareto \cite{PAR}, Mandelbrot
emphasized in the 1960s that
 fluctuations of financial assets
could be fitted much better
with the help of
L\'evy distributions\iems{L\'evy+distributions}\iems{L\'evy+distributions;,truncated}
\cite{MB1}.
These distributions
are defined by the Fourier transform
\begin{equation}
\tilde L^\mu_{ \sigma ^2}(x)\equiv
 \int _\infty^\infty\frac{dp}{2\pi}L^\mu_{ \sigma ^2}(p)\,e^{ipx},
\label{@interepL1}\end{equation}
with
\begin{equation}
 L^\mu_{ \sigma ^2}(p)\equiv
 \exp\left[- (\sigma^2p^2) ^{\mu/2}/2\right].
\label{@FTLEV}\end{equation}
The Gaussian distribution
are recovered in the limit
$\mu\rightarrow 2$.

For large $x$, the L\'evy distribution
(\ref{@interepL1})
falls off with the
characteristic power behavior
\begin{equation}
\tilde L^\mu_{ \sigma ^2}(x)\rightarrow A^\mu_{ \sigma ^2}\frac{\mu}{|x|^{1+\mu}}.
\label{@}\end{equation}
These power falloffs are referred to as {\em Paretian tails\/}\iems{Paretian tails}
of the distributions.
The amplitude of the tails is found by
approximating the integral (\ref{@interepL1}) for large $x$, where only small
momenta contribute, as follows
\begin{equation}
\tilde L^\mu_{ \sigma ^2}(x)\approx
 \int _{-\infty}^\infty\frac{dp}{2\pi}\left[1- \frac{1}{2}
(\sigma^2p^2) ^{\mu/2}\right]e^{ipx}
\mathop{\rightarrow}_{x\rightarrow \infty }A^\mu_{ \sigma ^2}\,\frac{\mu}{|x|^{1+\mu}},
\label{@interepL2}\end{equation}
with
\begin{equation}
A^\mu_{ \sigma ^2}=- \frac{ \sigma ^\mu}{2\mu}\int _0^\infty
\frac{dp'}{\pi}\,p'^\mu\cos p'
=\frac{ \sigma ^\mu}{2\pi \mu}\sin( \pi \mu/2)\, \Gamma (1+\mu).
\label{@}\end{equation}

The stock market data are fitted best with
 $\mu$ between $1.2$ and $1.5$ \cite{BP}, and we shall use
$\mu=3/2$ most of the time for simplicity,
where
one has
\begin{equation}
A^{3/2}_{ \sigma ^2}=\frac{1}{4} \frac{ \sigma ^{3/2}}{ \sqrt{ 2\pi}}.
\label{@}\end{equation}
The full Taylor expansion of the Fourier
transform
(\ref{@FTLEV}) yields the asymptotic  series
\begin{equation}
\tilde L^\mu_{ \sigma ^2}(x)=
  \sum_{n=0}^\infty
\frac{(-1)^n}{n!}\int _{0}^\infty\frac{dp}{\pi} \frac{
 \sigma^{\mu n}
  p^{\mu n}
}{2^n}\cos px=
 \sum_{n=0}^\infty
\frac{(-1)^{n+1}}{n!}\frac{ \sigma ^{\mu n}}{2^n\pi} \Gamma (1+n\mu)\sin(\pi\mu/2)\frac{1}{|x|^{1+\mu}}.
\label{@interepL}\end{equation}
This series  is not useful for practical
calculations since
 it fails to reproduce
 the exponential tails of the distribution,
a typical shortcoming of
large-$x$ expansions.
In particular,
it does not reduce to the Gaussian
distribution in the limit $\mu\rightarrow 2$.

\subsection{Truncated L\'evy Distributions}
\ins{truncated+L\'evy+distributions}\ins{distributions,L\'evy;truncated}%
\ins{L\'evy+distributions, truncated}
An undesirable property of the
L\'evy distributions  which is incompatible with financial data
is that their fluctuation width diverges
for $\mu<2$, since
\begin{equation}
 \sigma^2=\langle x^2 \rangle\equiv \int _{-\infty}^\infty dx\,x^2\,
\tilde L_{\sigma^2}^{\mu}(x)=\left.-\frac{d^2}{dp^2}
L_{ \sigma ^2}^\mu(p)\right|_{p=0}
\label{@quflw}\end{equation}
is infinite.
In contrast, real stock prices
  have a finite
width.
To account for both Paretian tails {\em and\/} finite
width one may introduce
the
so-called
 {\em truncated L\'evy distribution\/}s
\cite{MES}.\iems{L\'evy+distribution,truncated}
They are defined by
\begin{eqnarray}
\tilde L^{(\mu, \alpha) }_{ \sigma ^2}(x)\equiv
 \int _\infty^\infty\frac{dp}{2\pi}L^{(\mu, \alpha) }_{ \sigma ^2}(p)\,e^{ipx},
\label{@deftrld}\end{eqnarray}
with a
Fourier transform $L^{(\mu, \alpha) }_{ \sigma ^2}(p)$
 which generalizes the function (\ref{@FTLEV}).
It is
 conveniently
written as
an exponential
of a \lqq Hamiltonian
function\rqq{}   $H(p)$:
\begin{eqnarray}
 L_{\sigma^2}^{(\mu,\alpha)}(p)&\equiv &
e^{- H(p)},
\label{@LFFT0}\end{eqnarray}
with
\begin{eqnarray}
H(p)&\equiv &
\frac{ \sigma ^2}{2}
\frac{\alpha ^{2-\mu }}{\mu(1-\mu) }
\left[
 ( \alpha +ip)^ \mu+
 ( \alpha -ip)^ \mu-2 \alpha ^\mu
\right]
\nonumber \\
&=&
{ \sigma ^2}
\frac{( \alpha ^2+p^2)^{\mu/2}\cos[\mu\arctan(p/ \alpha )]- \alpha ^\mu}
{\alpha ^{\mu -2}\mu(1-\mu) }
.
\label{@LFFT}\end{eqnarray}

The asymptotic behavior of the
    truncated L\'evy distributions
differs from the power behavior
 of the
 L\'evy distribution
 in Eq.~(\ref{@interepL2})
by an exponential factor
$e^{- \alpha x}$, which guarantees the finiteness of the width $ \sigma $
and of
 all higher moments.
An estimate of the leading term is again obtained from the
Fourier transform
\begin{eqnarray}
\tilde L_{\sigma^2}^{(\mu,\alpha)} (x)
&\approx&
e^{2C  \alpha ^\mu}
 \int _{-\infty}^\infty\frac{dp}{2\pi}
\left\{1+C
\left[
 ( \alpha +ip)^ \mu+
 ( \alpha -ip)^ \mu
\right]
\right\}
e^{ipx}
\nonumber \\
&\displaystyle\mathop{\rightarrow}_{x\rightarrow \infty }&
e^{2C  \alpha ^\mu}
\Gamma (1+\mu)
\frac{\sin (\pi\mu)}{\pi}
C\frac{e^{- \alpha | x|}}
{|x|^{1+\mu}},
\label{@interepL2n}\end{eqnarray}
where
\begin{equation}
C\equiv\frac{ \sigma ^2}{2}
\frac{\alpha ^{2-\mu }}{\mu(1-\mu) }.
 \label{@}\end{equation}
The integral follows directly
from the formulas \cite{GR1}
\begin{equation}
\int _{-\infty }^\infty \frac{dp}{2\pi}\,( \alpha +ip)^{ \mu}
e^{ipx}=\frac{ \Theta(x)}{ \Gamma (- \mu)}\frac{e^{- \alpha x}}{x^{1+\mu}},
~~~~
\int _{-\infty }^\infty \frac{dp}{2\pi}\,( \alpha -ip)^{ \mu}
e^{ipx}=\frac{ \Theta(-x)}{ \Gamma (-\mu )}\frac{e^{- \alpha |x|}}{|x|^{1+\mu}},
\label{@rleq}\end{equation}
and the identity for Gamma functions
$1/ \Gamma (-z)=- \Gamma (1+z)\sin(\pi z)/\pi$.
The full expansion is integrated  with the help of the
formula   \cite{GR2}
\begin{eqnarray}
&&
\!\!\!\!\!\!
\!\!\!\!\!\!\!\!\!\!
\!\!\!\!\!\!\!\!\!\!
\int _{-\infty }^\infty \frac{dp}{2\pi}\,( \alpha +ip)^{ \mu }
( \alpha -ip)^{\nu}
e^{ipx}
\nonumber \\&&~~~\!\!\!\!\!\!\!\!\!\!
\!\!\!\!\!\!\!\!\!\!
=
{(2 \alpha) ^{\mu/2+ \nu/2 }}
\frac{1}{ |x|^{1+\mu/2+ \nu/2 }   }
%
\left\{    \dst
{\frac{1}{ \Gamma (- \mu )}
W_{(\nu- \mu )/2,(1+\mu+ \nu )/2}(2 \alpha x)  }
\atop    \dst
{\frac{1}{ \Gamma (- \nu )}
W_{(\mu- \nu )/2,(1+\mu+ \nu )/2}(2 \alpha x) }
\right.
~{\rm for }~{\raisebox{2mm}{$x>0$,} \atop \raisebox{-2mm}{$x<0$,}}
\label{@decomp}\end{eqnarray}
where
the Whittaker functions $W_{(\nu- \mu )/2,(1+\mu+ \nu )/2}(2 \alpha x) $
can be  expressed  in terms of
 Kummer's confluent  hypergeometric function
${}_1F_{1}(a;b;z)$
 as
\begin{eqnarray}
\!\!\!\!\!\!\!\!\!\!\!\!\!\!\!\!\!\!W_{ \lambda , \kappa }(z)&=&
\frac{  \Gamma (-2 \kappa )}{ \Gamma (1/2- \kappa - \lambda  )}
z^{ \kappa +1/2}e^{-z/2}{}_1F_{1}(1/2+ \kappa - \lambda ;2 \kappa +1;z)\nonumber \\
&+&\frac{  \Gamma (2 \kappa )}{ \Gamma (1/2+ \kappa - \lambda )}
z^{- \kappa +1/2}e^{-z/2}{}_1F_{1}( 1/2-\kappa - \lambda ;-2 \kappa +1;z).
\label{@}\end{eqnarray}
For $ \nu =0$, only  $x>0$ gives a nonzero integral (\ref{@decomp}), which
reduces, with
$W_{-\mu/2,1/2+\mu/2}(z)=z^{-\mu/2}e^{-z/2}$, to
the left equation in (\ref{@rleq}).
Setting $\mu= \nu $ we find
\begin{eqnarray}
&&
\!\!\!\!\!\!
\!\!\!\!\!\!\!\!\!\!
\!\!\!\!\!\!\!\!\!\!
\int _{-\infty }^\infty \frac{dp}{2\pi}\,( \alpha^2 +p^2)^{ \nu }
e^{ipx}
=
{(2 \alpha) ^{ \nu/2 }}
\frac{1}{ |x|^{1+ \nu }   }
\frac{1}{ \Gamma (- \nu )}
W_{0,1/2+ \nu }(2 \alpha| x|).
\label{@decomp1}\end{eqnarray}
Inserting
\begin{equation}
W_{0, 1/2+ \nu  }(z)= \sqrt{ \frac{2z}{\pi}}K_{1/2+  \nu   }(z/2),
\label{@}\end{equation}
we may write
\begin{eqnarray}
&&
\!\!\!\!\!\!
\!\!\!\!\!\!\!\!\!\!
\!\!\!\!\!\!\!\!\!\!
\int _{-\infty }^\infty \frac{dp}{2\pi}\,( \alpha^2 +p^2)^{ \nu }
e^{ipx}
=
{\left(\frac{2 \alpha }{|x|} \right)^{1/2+ \nu }}
\frac{1}{ \sqrt{\pi}  \Gamma (- \nu )}
K_{1/2+  \nu   }( \alpha |x|).
\label{@decomp2}\end{eqnarray}
For $ \nu =-1$ where
$K_{-1/2   }(z)
=K_{1/2   }(z)= \sqrt{\pi/2z}e^{-z}
$,
this
reduces          to
\begin{eqnarray}
&&
\!\!\!\!\!\!
\!\!\!\!\!\!\!\!\!\!
\!\!\!\!\!\!\!\!\!\!
\int _{-\infty }^\infty \frac{dp}{2\pi}\,\frac{1}{ \alpha^2 +p^2}
e^{ipx}
=
\frac{1}{2 \alpha }
e^{- \alpha |x|}.
\label{@decomp3}\end{eqnarray}
Summing up all terms
in the expansion
\begin{eqnarray}
\!\!\!\!\!\!\!\!\!\!\!\!\tilde L_{\sigma^2}^{(\mu,\alpha)} (x)
&\approx&
e^{2C \alpha ^\mu}
 \int _{-\infty}^\infty\frac{dp}{2\pi}
\left\{1+\sum _{n=1}^\infty
\frac{(-C)^n}{n!}
\left[
 ( \alpha +ip)^ \mu+
 ( \alpha -ip)^ \mu
\right]^n
\right\}
e^{ipx}
\label{@}\end{eqnarray}
yields the true asymptotic behavior
which differs  from
the estimate
(\ref{@interepL2n})
by a constant factor  \cite{KLy}:
\begin{eqnarray}
\tilde L_{\sigma^2}^{(\mu,\alpha)} (x)
&\displaystyle\mathop{\rightarrow}_{x\rightarrow \infty }&
e^{(2-2^\mu)C \alpha ^\mu}
\Gamma (1+\mu)
\frac{\sin (\pi\mu)}{\pi}
C\frac{e^{- \alpha | x|}}
{|x|^{1+\mu}}.
\label{@interepL2nn}\end{eqnarray}

\begin{figure}[tbhp]
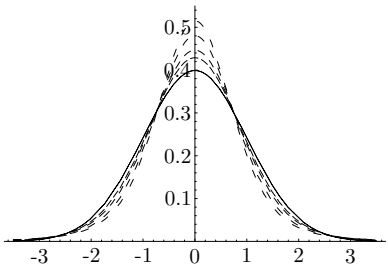

\unitlength .36mm
\input levys.tps \\
\caption[
Change in shape
of truncated L\'evy distributions
of width $ \sigma =1$ with increasing
kurtoses
$ \kappa =0$ (Gaussian, solid curve),$~1,\,2\,,5,\,10$
]
{
Change in shape
of truncated L\'evy distributions
of width $ \sigma =1$ with increasing
kurtoses
$ \kappa =0$ (Gaussian, solid curve),$~1,\,2\,,5,\,10$.
{} }
\label{@levys}\end{figure}

In contrast to  Gaussian distributions
which are
characterized completely by the width $ \sigma$,
the    truncated L\'evy distributions
contain
three parameters $ \sigma $, $\mu$, and $ \alpha $.
Best
fits to two types of fluctuating market prices are shown in Fig.~\ref{@markets},
in which we plot
the cumulative probabilities
\begin{eqnarray}
P_<( \delta x)=\int _{-\infty }^ {\delta x}dx\, \tilde L_{\sigma^2}^{(\mu,\alpha)} (x),~~~~~~
P_>( \delta x)=\int ^{\infty }_ {\delta x}dx\, \tilde L_{\sigma^2}^{(\mu,\alpha)} (x)=1-P_<( \delta x).
\label{@cumulD}\end{eqnarray}
For negative price fluctuations $ \delta x$,
the plot shows
$P_<( \delta x)$,
for positive price fluctuations
$P_>( \delta x)$.
By definition,
$P_<(-\infty )=0,\,
P_<(0)=1/2,\,
P_<(\infty )=1$, and
$P_>(-\infty )=1,\,
P_>(0)=1/2,\,
P_>(\infty )=0$.
To fit the general shape,
one chooses an appropriate parameter $\mu$ which turns out
to be rather universal, close to
$\mu=3/2$.
The remaining two parameters fix all expansion coefficients
of Hamiltonian (\ref{@LFFT}):
\begin{equation}
 H(p)=
\frac{1}{2} c_2\, p^2-\frac{1}{4!}c_4\,p^4
+\frac{1}{6!} c_6\, p^6-\frac{1}{8!}c_8\,p^8
+\dots~.
\label{@cumuatsH}\end{equation}
 The numbers $c_n$ are referred to as the {\em cumulants\/}%
\iems{cumulants,of+truncated+\'evy+distribution}%
\iems{L\'evy+distribution,truncated,cumulants}
of the truncated L\'evy distribution.
They are equal to
\begin{eqnarray}
c_2&=&{\sigma^2},
\nonumber \\
c_4&=&{\sigma^2}
(2-\mu)
(3-\mu)  \alpha ^{-2} ,
\nonumber \\
c_6&=&{\sigma^2}
(2-\mu)
(3-\mu)
(4-\mu)
(5-\mu)
\alpha ^{-4}           ,
\nonumber \\
&\vdots&\nonumber \\
c_{2n }&=& {\sigma^2}
\frac{ \Gamma (2n-\mu)}{ \Gamma (2-\mu)}
\alpha ^{2-2n},
\label{@}\end{eqnarray}
\begin{figure}[tbhp]
\unitlength .36mm
~\\[-.6cm]
$\!\!\!\!\!\!\!\!\!\!\!\!\!\!\!\!\!\!\!\!$
\input dmtousd.tps \\[3mm]
\caption[
Best fit of cumulative
versions (\protect\ref{@cumulD})
of truncated L\'evy
distribution,
$P_<( \delta x)$
for negative price fluctuations $ \delta x$,
and
$P_>( \delta x)$
for positive price fluctuations
to the instantaneous fluctuations of the S\&P 500 (l.h.s)
and the DM/US\$ exchange rate measured every fifteen minutes.
]{Best fit of cumulative
versions (\ref{@cumulD})
of truncated L\'evy
distribution,
$P_<( \delta x)$
for negative price fluctuations $ \delta x$,
and
$P_>( \delta x)$
for positive price fluctuations
to the instantaneous fluctuations of the S\&P 500 (l.h.s)
and the DM/US\$ exchange rate measured every fifteen minutes.
The  negative fluctuations lie on a  slightly higher curve than the positive
ones. The difference is often neglected.
The parameters $A$ and $ \alpha $ are the size and truncation parameters
of the distribution. The best value of $\mu$ is $3/2$
(from \cite{BP}).{}
}
\label{@markets}\end{figure}
The first determines the quadratic fluctuation width
\begin{equation}
\langle x^2 \rangle\equiv \int _{-\infty}^\infty dx\,x^2\,
\tilde L_{\sigma^2}^{(\mu,\alpha)} (x)=\left.-\frac{d^2}{dp^2}
L_{\sigma^2}^{(\mu,\alpha)}(p)\right|_{p=0}=
c_2= \sigma^2,
\label{@rulex2}\end{equation}
the second the fourth-order expectation
\begin{equation}
\langle x^4 \rangle\equiv \int _{-\infty}^\infty dx\,x^4\,
\tilde L_{\sigma^2}^{(\mu,\alpha)} (x)=\left.\frac{d^4}{dp^4}
L_{\sigma^2}^{(\mu,\alpha)}(p)\right|_{p=0}=c_4+3c_2^2,
\label{@rulex4}\end{equation}
and so on:
\begin{equation}
\langle x^6 \rangle
=c_6+15c_4c_2+15c_2^3,~~~~~~~
\langle x^8 \rangle
=c_8+28c_6c_2+35c_4^2+210c_4c_2^2+105c_2^4,\dots~.
\label{@rulexn}\end{equation}
In analyzing the data, one usually defines
the so-called {\em kurtosis\/}\iems{kurtosis},
which is the normalized fourth-order cumulant
\begin{equation}
 \kappa \equiv\bar c_4\equiv \frac{c_4}{c_2^2}=
 \frac{\langle x^4 \rangle}
{\langle x^2 \rangle^2}-3.
\label{@kappaform}\end{equation}
It depends on the parameters $ \sigma,\mu, \alpha $
as follows
\begin{equation}
 \kappa=\frac{(2-\mu)(3-\mu)}{ \sigma ^2 \alpha ^2
}.
\label{@kurtos}\end{equation}
Given the volatility $ \sigma $ and the kurtosis $ \kappa $,
we extract the L\'evy parameter $ \alpha $ from the equation
\begin{eqnarray}
\alpha =\frac{1}{ \sigma }\sqrt{\frac{(2-\mu)(3-\mu)}{ \kappa }}.~~~~~~
\label{Malp}\end{eqnarray}
In terms of $ \kappa $ and $ \sigma ^2$, the
expansion coefficients
are
\begin{eqnarray}
\bar c_4&=&  \kappa ,~~~
\bar c_6=  \kappa ^2\frac{(5-\mu)(4-\mu)}{(3-\mu)(2-\mu)},~~~
\bar c_8=  \kappa ^2\frac{
(7-\mu)(6-\mu)
(5-\mu)(4-\mu)
}{(3-\mu)^2(2-\mu)^2},~~~
\nonumber \\
\vdots\nonumber \\
\bar c_n&=&
 \kappa ^{n/2-1}
\frac{
 \Gamma (n-\mu) /\Gamma (4-\mu)
}{(3-\mu) ^{n/2-2}(2-\mu) ^{n/2-2}}.
\label{@coefbarcka}\end{eqnarray}
For $\mu=3/2$, the second equation in (\ref{Malp}) becomes simply
\begin{eqnarray}
 \alpha =\frac{1}{2} \sqrt{\frac{\,3}{  \sigma ^2 \kappa }}
 ,
\label{@}\end{eqnarray}
and
the coefficients
(\ref{@coefbarcka}):
\begin{eqnarray}
\bar c_4&=&  \kappa ,~~~
\bar c_6= \frac{5\cdot 7}{3} \,\kappa ^2,~~~
\bar c_8= 5\cdot 7\cdot 11\, \kappa ^2,~~~\nonumber \\
\vdots\nonumber \\
\bar c_{n}&=&
\frac{
 \Gamma (n-3/2) /\Gamma (5/2)
}{3 ^{n/2-2}/2^{n-4}}
 \kappa ^{n/2-1}
.~~~
\label{@coefbarcka}\end{eqnarray}
At zero kurtosis, the
\tld{} reduces to a Gaussian distribution
of width $ \sigma$.
The change in shape for a fixed width and increasing kurtosis
is shown in Fig.~\ref{@levys}.

From the S\&P and DM/US\$ data
 with time intervals $ \Delta t=15$\,min
one extracts $ \sigma^2=0.280 $ and $0.0163$,
and the kurtoses $ \kappa =12.7$ and $20.5$, respectively.
This implies
$M\approx3.57$, $ \alpha \approx0.46$  and
$M\approx61.35$,
$ \alpha \approx1.50$, respectively.

The other normalized cumulants
$(
\bar c_6,
\bar c_8,\dots)$
are then all determined
to be
$(1881.72, 788627.46,%
\dots)$
and $(-4902.92,3.3168\times 10^6$ $,%
\dots)$, respectively.
The cumulants increase rapidly showing that the expansion
needs resummation.

From the data, the other normalized cumulants are found by evaluating the
ratios of expectation values
\begin{eqnarray}
\bar c_6&=&
 \frac{\langle x^6 \rangle}
{\langle x^2 \rangle^3}
-15
 \frac{\langle x^4 \rangle}
{\langle x^2 \rangle^2}+30,\nonumber \\
\bar c_8&=&
 \frac{\langle x^8 \rangle}
{\langle x^2 \rangle^4}
 -28\frac{\langle x^6 \rangle}
{\langle x^2 \rangle^3}
 -35\frac{\langle x^4 \rangle^2}
{\langle x^2 \rangle^4}
+420
 \frac{\langle x^4 \rangle}
{\langle x^2 \rangle^2}-630,\dots~.
\label{@}\end{eqnarray}
%
%
%

\subsection{Asymmetric Truncated L\'evy Distributions}
\ins{truncated+L\'evy+distributions;asymmetric}\ins{distributions,L\'evy;asymmetric+truncated}%
\ins{L\'evy+distributions,truncated;asymmetric}%
\ins{asymmetric+truncated+L\'evy+distributions}%
We have seen in the data of Fig.~\ref{@markets}
that the price fluctuations
have a slight asymmetry: Price drops are slightly larger than
rises.
This is accounted for by an asymmetric
\tld{}.
It has the general form \cite{KOP}
\begin{eqnarray}
 L_{\sigma^2}^{(\lambda,\alpha, \beta )}(p)&\equiv &
e^{- H(p)},
\label{@LFFTHas}\end{eqnarray}
with a Hamiltonian
function
\begin{eqnarray}
\!\!\!\!\!\!\!\!\!\!\!\!\!\!\!\!\!\!\!\!\!\!\!H(p)&\equiv &
\frac{ \sigma ^2}{2}
\frac{\alpha ^{2-\lambda }}{\lambda(1-\lambda) }
\left[
 ( \alpha +ip)^ \lambda(1+ \beta )+
 ( \alpha -ip)^ \lambda(1- \beta )-2 \alpha ^\lambda
\right]
\nonumber \\
\!\!\!\!\!\!\!&=&
{ \sigma ^2}
\frac{( \alpha ^2+p^2)^{\lambda/2}
\left\{ \cos[\lambda\arctan(p/ \alpha )]+i  \beta
\sin[\lambda\arctan(p/ \alpha )]\right\}
- \alpha ^\lambda}
{\alpha ^{\lambda -2}\lambda(1-\lambda) }
.
\label{@LFFTas}\end{eqnarray}
This has
a power series
expansion
\begin{equation}
 H(p)=
ic_1p+\frac{1}{2} c_2\, p^2
-i\frac{1}{3!}c_3p^3-\frac{1}{4!}c_4\,p^4
+i\frac{1}{5!}c_5p^5
+\dots,
\label{@cumuatsHa}\end{equation}
which differs from
 (\ref{@cumuatsH})  by the extra odd coefficients:
\begin{eqnarray}
c_1&=& {\sigma^2} \frac{  \alpha  }{ (1-\lambda)} \beta   ,\\
c_3&=&
{\sigma^2} (2-\lambda){ \alpha^{-1} }  \beta  ,
\nonumber \\
c_5&=&
{\sigma^2}(2-\lambda)(3-\lambda)(4-\lambda){ \alpha^{-3} } \beta   ,
\nonumber \\
&\vdots,&\nonumber \\
c_{2n+1 }&=&  {\sigma^2} \frac{ \Gamma (2n+1-\lambda)}{ \Gamma (2-\lambda)}
\alpha ^{1-2n} \beta
,
\label{@}\end{eqnarray}
The general formula valid for even and odd $n$ is
\begin{eqnarray}
c_n= \sigma ^2\frac{ \Gamma (n-\lambda)}{ \Gamma (2-\lambda)}
\alpha ^{2-n}
\left\{ {1\atop  \beta} \right.~{\rm for}~~{\,n={\rm even},\atop
n={\rm odd.}}
\label{@}\end{eqnarray}
In addition to the even expectation values
(\ref{@rulex2})--(\ref{@rulexn})
 there are now        also odd expectation values:
\begin{eqnarray}
\langle x\rangle &\equiv &
 \int _{-\infty}^\infty dx\,x\,
\tilde L_{\sigma^2}^{(\lambda,\alpha, \beta )} (x)=\left.i\frac{d}{dp}
e^{-H(p)}\right|_{p=0}=
c_1,                \nonumber \\
\langle x^2\rangle &\equiv &
 \int _{-\infty}^\infty dx\,x^2\,
\tilde L_{\sigma^2}^{(\lambda,\alpha, \beta )} (x)=\left.-\frac{d^2}{d^2p}
e^{-H(p)}\right|_{p=0}=
c_2+c_1^2,                \nonumber \\
\langle x^3\rangle &\equiv &
 \int _{-\infty}^\infty dx\,x^3\,
\tilde L_{\sigma^2}^{(\lambda,\alpha, \beta )} (x)=\left.-i\frac{d^3}{d^3p}
e^{-H(p)}\right|_{p=0}=
c_3+3c_2c_1+c_1^3,                \nonumber \\
\langle x^4\rangle &\equiv &
 \int _{-\infty}^\infty dx\,x^4\,
\tilde L_{\sigma^2}^{(\lambda,\alpha, \beta )} (x)=\left.\frac{d^4}{d^4p}
e^{-H(p)}\right|_{p=0}=
c_4+4c_3c_1+3c_2^2+6c_2c_1^2+c_1^4,       \nonumber\\
~~~&\vdots&~~~. \label{@evenoddcfs}\end{eqnarray}
The distribution is now centered around a nonzero  average value:
\begin{equation}
\mu\equiv \langle x\rangle =c_1.
\label{@}\end{equation}
The fluctuation width is given by
\begin{equation}
 \sigma^2\equiv \langle x^2 \rangle- \langle x \rangle^2=
\Big\langle \left(x- \langle x \rangle\right)^2\Big\rangle=
c_2.
\label{@quflwn}\end{equation}

The asymptotic behavior of the
asymmetric  truncated L\'evy distributions
is given by a straightforward modification of (\ref{@rleq}):
\begin{eqnarray}
\tilde L_{\sigma^2}^{(\lambda,\alpha)} (x)
&\approx
&
 \int _{-\infty}^\infty\frac{dp}{2\pi}\,e^{ipx}\,
\left\{1\!- \!
\frac{ \sigma ^2}{2}
\frac{\alpha ^{2-\lambda }}{\lambda(1\!-\!\lambda) }
\left[
 ( \alpha +ip)^ \lambda(1\!+\! \beta )+
 ( \alpha -ip)^ \lambda(1\!- \!\beta )-2 \alpha ^\lambda
\right]
\right\}
\nonumber \\
&\displaystyle\mathop{\rightarrow}_{x\rightarrow \infty }&
~~ \sigma ^2   \,
e^{2s \alpha ^\lambda}
\Gamma (1+\lambda)
\frac{\sin (\pi\lambda)}{\pi}
s\frac{e^{- \alpha | x|}}
{|x|^{1+\lambda}}
[1+ \beta \,\sgn(x)].
\label{@interepL2nx}\end{eqnarray}
In analyzing the data, one now uses  the {\em skewness\/}\iems{skewness}
\begin{eqnarray}
s& \equiv&\bar c_3= \frac{c_3}{c_2^{3/2}}=
 \frac{1}{ \sigma ^3}\left(\langle x^3 \rangle-3c_2c_1-c_1^3\right)
=\frac{1}{ \sigma ^3}
\left(\langle x^3\rangle-3\Big\langle\left( x-\langle x\rangle\right)^2
\Big\rangle\langle x\rangle-\langle x\rangle ^3 \right)
\nonumber \\
&=&\frac{1}{ \sigma ^3}
\Big\langle\left( x-\langle x\rangle\right)^3\Big\rangle.
\label{@kappaform3}
\end{eqnarray}
 The kurtosis\ins{kurtosis} reads now
\begin{eqnarray}
 \kappa &\equiv&\bar c_4\equiv \frac{c_4}{c_2^2}=
 \frac{1}{ \sigma ^4}\left(
\langle x^4 \rangle-4c_3c_1-3c_2^2-6c_2c_1^2-c_1^4 \right) \nonumber \\
 &\!\!=\!\!& \frac{1}{ \sigma ^4}\left[
\langle x^4 \rangle-4
\Big\langle (x-\langle x\rangle )^3\Big\rangle \langle x\rangle
-3 \left(\langle x^2\rangle -\langle x\rangle^2 \right)^2
-6  \left(\langle x^2\rangle -\langle x\rangle^2 \right)\langle x\rangle^2
-\langle x\rangle^4 \right]          \nonumber \\
&\!\!=\!\!& \frac{1}{ \sigma ^4}\left[
\Big\langle (x-\langle x\rangle )^4\Big\rangle
-3\Big\langle x^2-\langle x\rangle^2 \Big\rangle^2
\right]                                           .
\label{@kappaform4}\end{eqnarray}
It depends on the parameters $ \sigma,\,\lambda,\, \beta  $, and $ \alpha $ or $ \kappa $
as follows
\begin{equation}
s=\frac{(2-\lambda) \beta }{ \sigma \alpha }.
\label{@skew}\end{equation}
The kurtosis has still the same dependence
(\ref{@kurtos}) on
$ \sigma,\, \lambda,\, \alpha $
as before, such that we may also write
\begin{equation}
s
= \sqrt{ \frac{(2-\lambda) \kappa }{(3-\lambda)  } }\beta .
\label{@skew2}\end{equation}
The average position
$\mu=\langle x\rangle $ is proportional to the ratio
of skewness and kurtosis:
\begin{equation}
\langle x\rangle =c_1=
 {\sigma^2} \frac{  \alpha  }{ (1-\lambda)} \beta
 =  \sigma \frac{3-\lambda}{1-\lambda}\frac{s}{ \kappa }.
\label{@}\end{equation}
From the data one extracts
 volatility $ \sigma $, kurtosis $ \kappa $,
and skewness $s$
which
determine
completely the asymmetric truncated   L\'evy distribution.
The data are then plotted
against $x-\langle x\rangle =x-\mu$, so that they
 are centered
 at the average position. This centered
distribution will be denoted by $\bar
 L_{\sigma^2}^{(\lambda,\alpha, \beta )} (x)$, i.e.
\begin{equation}
 \bar
 L_{\sigma^2}^{(\lambda,\alpha, \beta )} (x)\equiv
\tilde
 L_{\sigma^2}^{(\lambda,\alpha, \beta )} (x-\mu)
\label{@}\end{equation}
The Hamiltonian associated with this
zero-average distribution is
\begin{equation}
\bar H(p)\equiv
\bar H(p)-H'(0)p,
\label{@20batH}\end{equation}
and its expansion in power of the momenta starts out with $p^2$, i.e., the first term in
 (\ref{@cumuatsHa}) is subtracted.

\begin{figure}[tbhp]
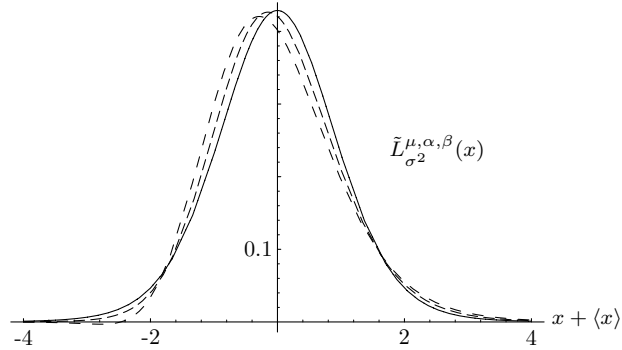

\unitlength .36mm
\input levysa.tps \\[3mm]
\caption[
Change in shape
of truncated L\'evy distributions
of width $ \sigma =1$ and kurtosis
$ \kappa =1$ with increasing
skewness
$ s=0$ (solid curve),~$0.4,\,0.8$
]
{
Change in shape
of truncated L\'evy distributions
of width $ \sigma =1$ and kurtosis
$ \kappa =1$ with increasing
skewness
$ s=0$ (solid curve),~$0.4,\,0.8\,$.
The curves are centered around $\langle x\rangle $.
{} }
\label{@levysa}\end{figure}

In terms of $ \kappa $, $ \sigma $, and $s$, the
normalized expansion coefficients
are
\begin{eqnarray}
\!\!\!\!\!\!\!\!\!\!\!\!\!\!\!\!\!\!\!\bar c_n&
=&
 \kappa ^{n/2-1}
\frac{
 \Gamma (n-\lambda) /\Gamma (4-\lambda)
}{(3-\lambda) ^{n/2-2}(2-\lambda) ^{n/2-2}}
\left\{ {\raisebox{2mm}{1}\atop  \raisebox{-2mm}{$\sqrt{(3-\lambda)/(2-\lambda) \kappa }\,s$} }  \right.~{\rm for}~
~~{\raisebox{2mm}{$n={\rm even},$}\atop
\raisebox{-2mm}{$n={\rm odd.}$}}
\label{@coefbarckas}\end{eqnarray}
The change in shape of the distributions
of a fixed width and kurtosis
with increasing skewness
is shown in Fig.~\ref{@levysa}.
We have plotted the distributions
centered around the average position
$x=\langle c_1\rangle $
which means that we have removed from $H(p)$
in (\ref{@LFFTHas}),
(\ref{@LFFTas}), and (\ref{@cumuatsHa})
the linear term $ic_1p$.
This subtracted Hamiltonian whose power series
expansion begins with the term $c_2p^2/2$ will be denoted by
\begin{equation}
\bar H(p)\equiv
H(p)-H'(0)=
\frac{1}{2} c_2\, p^2
-i\frac{1}{3!}c_3p^3-\frac{1}{4!}c_4\,p^4
+i\frac{1}{5!}c_5p^5
+\dots.
\label{@}\end{equation}

\subsection{Meixner Distributions}
Quite reasonable
 fits to financial data
are provided by the  \iem{Meixner distributions} \cite{Gri,Mei}.
They have the advantage that
they can be given explicitly in
configuration and momentum space
\begin{eqnarray}
\tilde M(x)&=&
\frac{\left[2\cos (b/2)\right]^{2d}}{2a\pi \Gamma (2d)}
| \Gamma \left(d+ix/a\right) |^2\exp\left[bx/a\right] , \\
M(p)&=& \left\{\frac{\cos(b/2)}{\cosh\left[(ap-ib)/2 \right] } \right\} ^{2d}
.
\label{@}\end{eqnarray}
They have the same tail behavior as the \tld{}s
\begin{eqnarray}
\tilde M(x)\rightarrow C_\pm|x|^ \rho e^{- \sigma_{\pm} |x|}~~~~{\rm for}~~~x\rightarrow \pm \infty ,
\label{@}\end{eqnarray}
with
\begin{eqnarray}
C_\pm=
\frac{\left[2\cos (b/2)\right]^{2d}}{2a\pi \Gamma (2d)}
 \frac{2\pi}{a^{2d-1}}e^{\pm 2\pi d\tan(b/2)},~
 \rho =2d-1,~\sigma_\pm\equiv (\pi\pm b)/a.\label{@}\end{eqnarray}
The moments are
\begin{eqnarray}
\mu&\!\!\!=\!\!\!&ad\tan (b/2),~~~~\,~~~~~~~~~~~~
 \sigma ^2=a^2d/2\cos^2 (b/2),~~   \nonumber \\
s&\!\!\!=\!\!\!&- \sqrt{2} \sin(b/2)/ \sqrt{d},~~~
 ~~~~~~\hspace{1pt} \kappa =[2-\cos b]/d,
\label{@}\end{eqnarray}
such that we can calculate the parameters
from the moments as follows:
\begin{eqnarray}
\!\!\!\!\!\!a^2= \sigma ^2\left(2 \kappa -3s^2\right),~~~
d=\frac{1}{ \kappa -s^2},~~~
b=-2\arcsin \left(s\sqrt{  d/2}\right).
\label{@}\end{eqnarray}

\begin{figure}[tbhp]
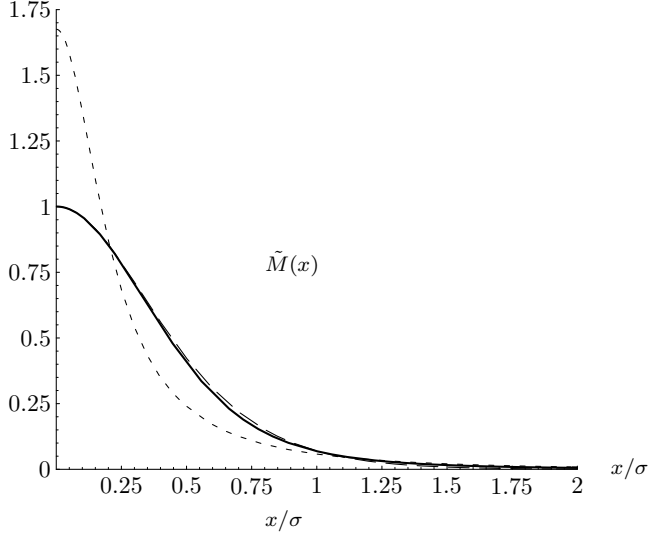

\input meixlev.tps
~\\ [3mm]
\caption[Comparison of best fit of Meixner distribution
to \tld{}]{Comparison of best fit of Meixner distribution
to \tld{} one by using the same
$ \sigma ^2$ and the kurtosis $ \kappa $ (short dashed)
and once using the same
value and curvature at the origin (long-dashed).
The parameters are $ \sigma ^2=0.280$ and $ \kappa =12.7$
as in the left-hand cumulative distribution
in Fig.~\ref{@markets}. The  Meixner distribution
with the same $ \sigma ^2$ and $ \kappa $ has
parameters $a=2.666,\,d=0.079,\,b=0$, the distribution with the same
value and curvature at the origin has
 $a=0.6145,\,d=1.059,\,b=0$.
The very large $ \sigma $-regime, however, is not fitted well as can be seen
in the cumulative distributions
which reach out to $x$ of the order of $10 \sigma $ in Fig.~\ref{@markets}.}
\label{compMeiLe}\end{figure}

The Meixner distributions
can be fitted quite well to  the \tld{} in the regime of large
probability.
In doing so we observe that the variance
$ \sigma ^2$ and the kurtosis $ \kappa $
are not the best parameters
to match the two distributions.
The large-probability
regime of the distributions can be matched perfectly
by choosing, in the symmetric case,
 the value
 and the curvature at the origin to be the same in both curves.
This is seen in Fig.~\ref{compMeiLe}.

In the asymmetric case
we have to match also the first and third
derivatives.
The derivatives
of the Meixner distribution are:
\begin{eqnarray}
\tilde M(0)&=&\frac{2^{2d-1} \Gamma^2 (d)}{\pi a \Gamma (2d)},\nonumber \\
\tilde M'(0)&=&b\frac{2^{2d-1} \Gamma^2 (d)\left[ 1-d\psi(d)\right] }{\pi a^2 \Gamma (2d)},\nonumber \\
\tilde M''(0)&=&-\frac{2^{2d} \Gamma^2 (d)\psi(d)}{\pi a^3 \Gamma (2d)},\nonumber \\
\tilde M^{(3)}(0)&=&-\frac{b}{2}\,\frac{2^{2d} \Gamma^2 (d)\left[6\psi(d)
-6d\psi^2(d)-d\psi^{(3)}(d)\right]}{\pi a^4 \Gamma (2d)},\nonumber \\
\tilde M^{(4)}(0)&=&\frac{2^{2d} \Gamma^2 (d)\left[6\psi^2(d)
+\psi^{(3)}(d)\right] }{\pi a^5 \Gamma (2d)},
\label{@20slopesM}\end{eqnarray}
where $\psi^{(n)}(z)\equiv d^{n+1}\log  \Gamma (z)/dz^{n+1}$ are the Polygamma functions.

\subsection{Other Non-Gaussian  Distributions}

As mentioned in the beginning,
some processes are also fitted
quite well be generalized hyperbolic distributions
\cite{N1,N2,N2a,N3,N4,N5,N6,N7,N8,N9,N9b,N10,N11,N12,N13,N14,N15,N16,N17,N18,N19,N20,N21,N22,N23}.
They
read
\begin{eqnarray}
\!\!\!\!\! \tilde H_G(x)=
\frac{\,
    {\left( {\gamma}^2\! - {\beta}^2 \right)^{ \lambda /2}
e^{\beta\, x } }    }{{\gamma}^{ \lambda -\lfrac{1}{2} }
{\delta}^
     {\lambda}\,{\sqrt{2\,\pi }}\,
   }
\,
    {\left[ {\delta}^2 \!+ x^2 \right] }^
     { \lambda /2-1/4}
    \frac{K_{ \lambda -1/2}\left(
     \gamma\,{\sqrt{{\delta}^2 \!+
          x^2}}\right)}{ K_\lambda\left(
     { \delta \sqrt{ {\gamma}^2 - {\beta}^2}}
\right)}
\label{@GHFUNCTIONS}\end{eqnarray}
and
\begin{eqnarray}
G(p)=
\frac{\delta ^ \lambda
\left( {\gamma}^2 - {\beta}^2 \right)^{ \lambda /2}}
{K_ \lambda \left(  \delta
     {\sqrt{ {\gamma}^2 - {\beta}^2}}
 \right)
 }{
   }
\,
    \frac{K_{ \lambda}\left(
      \delta {\sqrt{ \gamma ^2 -
          {\left( \beta +ip\right) }^2}}
\right)}{
\left[      \delta {\sqrt{ \gamma ^2 -
          {\left( \beta +ip\right) }^2}}\right]^ \lambda
}  ,
\label{@20HsubG}\end{eqnarray}
the latter defining another Hamiltonian
\begin{equation}
H_G(p)\equiv -\log G(p).
\label{@}\end{equation}
Introducing the variable
$ \zeta \equiv  \delta  \sqrt{{ \gamma ^2- \beta ^2}}$, this can be expanded
in powers of $p$ as in Eq.~(\ref{@cumuatsH}), yielding
 the first two cumulants:
\begin{eqnarray}
c_1&=& \beta \frac{ \delta ^2} \zeta
\frac{K_{1+ \lambda }( \zeta )}{K_{ \lambda }( \zeta )};
\\
c_2&=&  \frac{ \delta ^2} \zeta
\frac{K_{1+ \lambda }( \zeta )}{K_{\lambda }( \zeta )}
+ \frac{\beta ^2 \delta ^4}
 {\zeta^2}
\left\{
\frac{K_{2+ \lambda }( \zeta )}{K_{ \lambda }( \zeta )}
-\left[
\frac{K_{1+ \lambda }( \zeta )}{K_{ \lambda }( \zeta )}
\right] ^2
\right\}.
\label{@}\end{eqnarray}
Using the identity
\begin{equation}
K_{ \nu+1}(z)-
K_{ \nu-1}(z)=\frac{2 \nu }{z}
K_{ \nu}(z),
\label{@}\end{equation}
the latter equation can be expressed entirely in terms of
\begin{eqnarray}
r=r( \zeta)=  \frac{K_{1+ \lambda }( \zeta )}{K_{ \lambda }( \zeta )}
\label{@rfunction1}\end{eqnarray}
as
\begin{eqnarray}
c_2&=&
  \frac{ \delta ^2} \zeta
\frac{K_{1+ \lambda }( \zeta )}{K_{\lambda }( \zeta )}
+ \frac{\beta ^2 \delta ^4}
 {\zeta^3}
\left\{   \zeta +2\left(1+ \lambda \right)
\frac{K_{1+ \lambda }( \zeta )}{K_{ \lambda }( \zeta )}
 -\zeta \left[
\frac{K_{1+ \lambda }( \zeta )}{K_{ \lambda }( \zeta )}
\right] ^2
\right\}
\nonumber \\&=&
  \frac{ \delta ^2} \zeta r
+ \frac{\beta ^2 \delta ^4}
 {\zeta^3}
\left[   \zeta +2\left(1+ \lambda \right)
r
 -\zeta
r^2
\right]
.
\label{@}\end{eqnarray}
Usually,  the asymmetry of the distribution is small
implying that $c_1$ and thus $ \beta $ is small.
It is then useful to introduce the
symmetric variance
\begin{eqnarray}
\sigma _s^2\equiv  \delta ^2r/ \zeta,
\label{@sigmas}\end{eqnarray}
and write
\begin{equation}
c_1= \beta  \sigma _s,~~~c_2= \sigma ^2= \sigma _s^2 +
 \beta ^2\left[
\frac{ \delta ^4}{ \zeta ^2}
+2\left(1+ \lambda \right)\frac{ \delta ^2}{ \zeta ^2} \sigma _s- \sigma _s^2
\right]
\label{@sigmavonsigmas}\end{equation}
The cumulants $c_3$ and $c_4$ are most compactly written
 as
\begin{eqnarray}
\!\!\!\!\!\!\!\!\!\!\!c_3&=& \beta
\left[ \frac{3 \delta ^4}{ \zeta ^2}+6\left(1+ \lambda \right)
\frac{ \delta ^2}{ \zeta ^2}  \sigma _s^2-3 \sigma _s^4\right]
\nonumber \\
\!\!\!\!\!\!\!\!\!\!\!\!\!&+ &\beta ^3
\left\{ 2\left(2+ \lambda \right)\frac{ \delta ^6}{ \zeta ^4}
+ \left[4\left(1+ \lambda\right)\left(2+ \lambda \right)-
2 \zeta^2
 \right]\frac{ \delta ^4}{ \zeta ^4} \sigma _s^2
-6\left(1+ \lambda \right)
\frac{ \delta ^2}{ \zeta ^2}~\sigma _s^4+2 \sigma _s^6\right\}~~~~
\label{@c3equatio}\end{eqnarray}
and
\begin{eqnarray}
\!\!\!\!\!\!\!\!\!\!\!c_4&=&  \kappa  \sigma ^4=
 \frac{3 \delta ^4}{ \zeta ^2}
+
\frac{ 6\delta ^2}{ \zeta ^2} \left(1+ \lambda \right)
 \sigma _s^2-3 \sigma _s^4
\nonumber \\
\!\!\!\!\!\!\!\!\!\!\!\!\!&+ &6\beta ^2
\left\{ 2\left(2+ \lambda \right)\frac{ \delta ^6}{ \zeta ^4}
+ \left[4\left(1+ \lambda\right)\left(2+ \lambda \right)-2 \zeta^2
 \right]\frac{ \delta ^4}{ \zeta ^4} \sigma _s^2
-6\left(1+ \lambda \right)
\frac{ \delta ^2}{ \zeta ^2}~\sigma _s^4+2 \sigma _s^6\right\}~~~~
\nonumber \\
\!\!\!\!\!\!\!\!\!\!\!\!\!&+ &~~\!\beta ^4
\left\{ \left[ 4\left(2+ \lambda \right)\left(3+ \lambda \right) -
\zeta ^2
\right]
\frac{ \delta ^8}{ \zeta ^6}
+ \left[4\left(1+ \lambda\right)\left(2+ \lambda \right)
\left(3+ \lambda \right)
-2\left(5+4 \lambda \right) \zeta^2
 \right]\frac{ \delta ^6}{ \zeta ^6} \sigma _s^2   \right.
\nonumber \\
&\!\! &\left.~~~~~-2   \left[ \left(1+ \lambda \right)\left(11+7 \lambda \right)
-2 \zeta ^2\right] \frac{ \delta ^4}{ \zeta ^4}
 \sigma _s^4
+12
\left(1+ \lambda \right)\frac{ \delta ^2}{ \zeta ^2} \sigma _s^6
-3 \sigma _s^8
\right\}.~~~~
\label{@c4equation}\end{eqnarray}
The first term in $c_4$ is equal to $\sigma _s^4$
times
 the kurtosis of the symmetric distribution
\begin{eqnarray}
 \kappa _s\equiv
 \frac{3 \delta ^4}{ \zeta ^2\sigma _s^4}
+
\frac{ 6\delta ^2}{ \zeta ^2\sigma _s^2} \left(1+ \lambda \right)
 -3,
\label{@}\end{eqnarray}
Inserting here $ \sigma _s^2$ from
(\ref{@sigmas}), we find
\begin{eqnarray}
 \kappa _s\equiv\frac{3}{r^2( \zeta )}+(1+ \lambda )
\frac{6}{ \zeta \,r( \zeta )}-3.
\label{@kappsofl}\end{eqnarray}
Since all Bessel functions $K_ \nu (z)$ have the same large-$z$
behavior $K_ \nu (z)\rightarrow  \sqrt{\pi/2z}e^{-z}$
and the small-$z$ behavior $K_ \nu (z)\rightarrow  \Gamma ( \nu )/2(z/2)^ \nu $,
the kurtosis
starts out at $3/ \lambda $ for $ \zeta =0$ and decreases monotonously
to 0 for $ \zeta \rightarrow \infty $.
Thus a high kurtosis can be reached only with a small
parameter $ \lambda $.

The first term in $c_3$ is
$ \beta  \kappa _s \sigma _s^4$, and the first two terms in $c_4$ are
$ \kappa _s \sigma _s^4+6\left(c_3/ \beta - \kappa _s \sigma _s^4\right)$.
For a symmetric distribution
with  certain variance
$ \sigma _s^2$ and kurtosis
$ \kappa _s$
we
 select some
parameter $ \lambda<3/ \kappa _s$,
and
solve the Eq.~(\ref{@kappsofl})
to find $ \zeta $.
This is inserted into
Eq.~(\ref{@sigmas})
to determine
\begin{eqnarray}
 \delta ^2&=&\frac{ \sigma _s^2 \,\zeta }{r( \zeta )}.
\label{@delatsqueq}\end{eqnarray}
For larger kurtosis,
the kurtosis is not an optimal
parameter to determine generalized hyperbolic distributions.
A better fit to the data
is reached by
reproducing correctly
 the size and shape of the
 near the peak
and allow for some deviations
in the tails of the distribution, on which
the kurtosis depends quite sensitively.

For
distributions which are only
 slightly asymmetric, which is usually
the case,
it is  sufficient
to solve the above symmetric equations
and determine the  small parameter $ \beta $ approximately
by the skew   $s=c_3/ \sigma ^3$
from the first line in (\ref{@c3equatio})
as
\begin{equation}
 \beta \approx\frac{s}{ \kappa _s \sigma _s}.
\label{@}\end{equation}
This approximation
can be improved iteratively
by reinserting $ \beta $  into
the second equation in (\ref{@sigmavonsigmas})
to determine from the variance of the data
$ \sigma ^2$
an improved value of $ \sigma  _s^2$,
then into the first
two lines of  Eq.~(\ref{@c4equation})
to determine from the kurtosis $ \kappa $ of the data
an improved value of $ \kappa _s$,
and so on.

\subsection{Debye-Waller Factor for Non-Gaussian Fluctuations}
An important quantity for fluctuating Gaussian variables
is the \iems{Debye-Waller factor}.
It was introduced in solid state physics
to describe the reduction
of intensity of Bragg peaks due to the thermal
fluctuations
of the atomic positions.
If ${\bf u}({\bf x})$ is the atomic displacement field,
 the
Gaussian approximation to the
Debye-Waller factor $e^{-2W}$
is given by
\begin{equation}
e^{-W}\equiv
\langle e^{-\nablabs\cdot\sbf u(\sbf x)}\rangle
=e^{-\Sigma_{\ssbf k}\langle\,| {\sbf k}\cdot \sbf u(\sbf k)]^2\rangle /2}
,
\label{@}\end{equation}
This is
a
 direct consequence
of the relation for Gauss distributions
\begin{equation}
\langle e^{Px}\rangle\equiv \int dx\,
\frac{1}{\sqrt{2\pi \sigma^2}}
e^{-x^2/2\sigma^2}
e^{Px}
=\int dx\, \int \frac{dp}{2\pi}\,e^{-\sigma^2p^2/2}
e^{ipx+Px}
=e^{\sigma^2P^2/2},
\label{@modwick2}\end{equation}
which is a manifestation of
Wick's rule
\begin{equation}
\langle e^{Px}\rangle=
e^{P^2\langle x^2\rangle/2},
\label{@}\end{equation}
whose  right-hand side may be considered
as
 the
Debye-Waller factor for the Gaussian variable $x$.

There exists a simple generalization of this  relation
to  non-Gaussian  distributions, which is
\begin{equation}
\langle e^{Px}\rangle\equiv \int dx\, \tilde L_{ \sigma ^2}^{(\mu, \alpha) }(x)
e^{Px}
=\int dx\, \int \frac{dp}{2\pi}\,e^{-H(p)}
e^{ipx+Px}
=e^{-H(iP)}.
\label{@modwick}\end{equation}

\subsection{Path Integral for Non-Gaussian Distribution}

Let us calculate the properties of the
simplest process whose fluctuations
are distributed according to any of the general
non-Gaussian distributions.
We consider the stochastic differential equation
\begin{equation}
\dot x(t)= r_{x}+\eta (t),
\label{@stdeqet}\end{equation}
where the noise variable $ \eta (t)$ is distributed according
an arbitrary distribution.
The constant growth rate $r_{x}$ in (\ref{@stdeqet})
is uniquely defined only if the average of the  noise variable vanishes: $ \langle \eta (t)\rangle =0$.
The general distributions
discussed above can have a nonzero average
$\langle x\rangle =c_1$ which has to be subtracted
from $ \eta (t)$ to
identify $r_{x}$.
The
subsequent discussion
will be simplest if we imagine
$r_{x}$ to have replaced $c_1$ in the above distributions,
i.e., that the power series expansion of the Hamiltonian
(\ref{@cumuatsHa}) is replaces as follows:
\begin{eqnarray}
\!\!\!\!\!\!\!\!\!\!\!\! H(p)\rightarrow H_{r_{x}}(p)&\equiv&
H(p)-H'(0)p+ir_{x}p\equiv
\bar H(p)+ir_{x}p
\nonumber \\
&\equiv &
ir_{x}\,p+\frac{1}{2} c_2\, p^2
-i\frac{1}{3!}c_3p^3-\frac{1}{4!}c_4\,p^4
+i\frac{1}{5!}c_5p^5
+\dots~.
\label{@cumuatsHar}\end{eqnarray}
Thus we may simply work with the original
expansion (\ref{@cumuatsHa})
and replace, at the end,
\begin{equation}
c_1\rightarrow r_{x}.
\label{@c1modif}\end{equation}
The stochastic differential equation
(\ref{@stdeqet})
can be assumed to read
simply
\begin{equation}
\dot x(t)=\eta (t),
\label{@stdeqet2}\end{equation}
With the ultimate
replacement
(\ref{@c1modif}) in mind,
the  probability distribution of the endpoints
for the paths starting at a certain initial
point is given by a path integral of the form
%
\begin{eqnarray}
P(x_b t_b|x_at_a)=
\int {{\cal D}  \eta }
\int {{\cal D}  x}
\exp\left[ -\int _{t_a}^{t_b}dt \, \tilde H( \eta (t))\right]
 \delta [\dot x-  \eta ],
\label{@18spiLP}
\end{eqnarray}
with the initial condition
$x(t_a)=x_a$. The final point is, of course, $x_b=x(t_b)$.

The function $\tilde H( \eta )$ to be modified at the end
by the replacement
(\ref{@cumuatsHar})
 is
the negative logarithm of the \tld{} or the generalized hyperbolic distribution
or any other  distribution. By analogy with the definition in
momentum space
(\ref{@LFFT0})
we shall define $\tilde H( \eta )$ by
\begin{equation}
e^{-\tilde H(x)}\equiv  \tilde L_{ \sigma ^2}^{(\lambda, \alpha )}(x),~~~~
{\rm or}~~~~
e^{-\tilde H(x)}\equiv  \tilde G(x),
\label{@}\end{equation}
or for any
distribution function $\tilde D(x)$ generically by
\begin{eqnarray}
      e^{-\tilde H(x)}
\equiv \tilde D(x).
\label{@20gendisD}\end{eqnarray}

The correlation functions of the noise
variable $\eta(t)$ in the path integral
(\ref{@18spiLP})
are given by a straightforward functional generalization
of formulas
(\ref{@evenoddcfs}).
For this purpose,
we
 express the noise distribution
 $P[ \eta ]\equiv \exp\left[ {-\int_{t_a}^{t_b} dt\,\tilde H(\eta(t) )}\right] $
in
(\ref{@18spiLP})
as a Fourier path integral
\begin{eqnarray}\!\!\!\!\!\!\!\!\!
P[ \eta]= \int {{\cal D}  \eta } \!
\int \frac{{\cal D}  p }{2\pi}
\exp\left\{
\int _{t_a}^{t_b}\!dt\left[ip(t) \eta (t)-  H( p(t))\right] \right\},
\label{@18spiLP1n}
\end{eqnarray}
and note that the correlation functions
can be obtained from
the functional derivatives
\begin{eqnarray}
&&\!\!\!\!\!\!\!\!\!\!\!\!\!\!\!
\langle  \eta (t_1)\cdots \eta (t_n)\rangle =(-i)^n
\int {{\cal D}  \eta } \!
 \int \frac{{\cal D}  p }{2\pi}
\left[\frac{ \delta }{ \delta p(t_1)}                    \cdots
\frac{ \delta }{ \delta p(t_n)}
e^{i \int _{t_a}^{t_b}\!dt\,p(t) \eta (t)} \right]
e^{- \int _{t_a}^{t_b}\!dt\,  H( p(t)) }. \nonumber
\label{@}\end{eqnarray}
After $n$  partial integrations, this becomes
\begin{eqnarray}
\!\!\!\!\!\!\!\!\!\!\!\!
\langle  \eta (t_1)\cdots \eta (t_n)\rangle &=&
i^n
\int {{\cal D}  \eta } \!
\int \frac{{\cal D}  p }{2\pi}  e^{i \int _{t_a}^{t_b}\!dt\,p(t) \eta (t)}
\, \frac{ \delta }{ \delta p(t_1)}                    \cdots
\frac{ \delta }{ \delta p(t_n)}
 e^{- \int _{t_a}^{t_b}\!dt\,  H( p(t)) }
       ~~~~~~\nonumber \\[2mm]
&=&i^n
\left[ \frac{ \delta }{ \delta p(t_1)}                    \cdots
\frac{ \delta }{ \delta p(t_n)}
 e^{- \int _{t_a}^{t_b}\!dt\,  H( p(t)) }
    \right]_{p(t)\equiv 0} . ~~~~~~~~~
\label{@18spiLP1n1}
\end{eqnarray}
By expanding the exponential  $
 e^{- \int _{t_a}^{t_b}\!dt\,  H( p(t)) }$ in a power series
 (\ref{@cumuatsHa})  we find
immediately
\begin{eqnarray}
\langle  \eta (t_1)  \rangle
&\equiv& Z^{-1}
\int {{\cal D}  \eta } \,\eta (t_1)
\exp\left[ -\int _{t_a}^{t_b}dt \, \tilde H( \eta (t))\right]
=c_1,\label{@correlfu1} \\
\langle  \eta (t_1) \eta (t_2) \rangle
&\equiv&   Z^{-1}
\int {{\cal D}  \eta } \,\eta (t_1) \eta (t_2)
\exp\left[ -\int _{t_a}^{t_b}dt \, \tilde H( \eta (t))\right]
\nonumber \\
&=&c_2 \delta (t_1-t_2)+c_1^2,
\label{@correlfu2}\\
\langle
 \eta (t_1) \eta (t_2) \eta (t_3)
 \rangle
&\equiv& Z^{-1}
\int {{\cal D}  \eta }\,
 \eta (t_1) \eta (t_2) \eta (t_3)
\exp\left[ -\int _{t_a}^{t_b}dt \, \tilde H( \eta (t))\right]
\nonumber \\
&=&
c_3
\delta (t_1\!-\!t_2)
\delta (t_1\!-\!t_3)
  \\
\!\!\!&\!\!\!\!\!\!+\!\!\!\!\!\!&c_2c_1
\big[
\delta (t_1\!-\!t_2)+\delta (t_2\!-\!t_3)\!+\!
\delta (t_1\!-\!t_3)
\big]+c_1^3 ,
\nonumber \\
\langle
 \eta (t_1) \eta (t_2) \eta (t_3) \eta (t_4)
 \rangle
&\equiv&  Z^{-1}
\int {{\cal D}  \eta }\,
 \eta (t_1) \eta (t_2) \eta (t_3) \eta (t_4)
\exp\left[ -\int _{t_a}^{t_b}dt \, \tilde H( \eta (t))\right]
\nonumber \\\!\!\!&=&
c_4
\delta (t_1\!-\!t_2)
\delta (t_1\!-\!t_3)
\delta (t_1\!-\!t_4)
\nonumber \\\!\!\!&\!\!\!+\!\!\!&
c_3c_1
\big[
\delta (t_1\!-\!t_2)\delta (t_1\!-\!t_3)
+3\,{\rm cyclic~perms}
\big]
 \label{@correlfu} \\
\!\!\!&\!\!\!+\!\!\!&
c_2^2
\big[
\delta (t_1\!-\!t_2)\delta (t_3\!-\!t_4)+
\delta (t_1\!-\!t_3)\delta (t_2\!-\!t_4) +
\delta (t_1\!-\!t_4)\delta (t_2\!-\!t_3)\big]\nonumber \\\nonumber
\!\!\!&\!\!\!\!\!\!+\!\!\!\!\!\!&c_2c_1^2
\big[
\delta (t_1\!-\!t_2)+5\,{\rm pair~terms}\big]+c_1^4,
\end{eqnarray}
where
\begin{eqnarray}
Z\equiv
\int {{\cal D}  \eta } \,\exp\left[ -\int _{t_a}^{t_b}dt \, \tilde H( \eta (t))\right].
 \label{@}\end{eqnarray}
The higher
 correlation functions are obvious generalizations of
(\ref{@rulexn}).
The different contributions on the right-hand side of
(\ref{@correlfu2})--(\ref{@correlfu})
 Eq.~(\ref{@correlfu}) are distinguishable
by their connectedness structure.

An important property of the probability
(\ref{@18spiLP}) is that it satisfies
the semigroup property
\begin{eqnarray}
P(x_c t_c|x_at_a)=
\int_{-\infty}^\infty
 dx_b\,P(x_c t_c|x_bt_b)P(x_b t_b|x_at_a).
\label{@convprtop}\end{eqnarray}
In  Fig.~\ref{@convsp} we show that
  this property is satisfied by
experimental
asset distributions
reasonably well, except in  the low-probability tails.
The discrepancy manifests itself also at another place:
We shall see in
Subsection
\ref{@FPLEQ} that
the solution of the path integral has a kurtosis decreasing
inversely proportional
to the time. The data
 in Fig.~\ref{@convsp}, however, show only an slower inverse square-root
falloff.
This can be accounted for
in the theory by including
fluctuations of the  width  $ \sigma $,
which are certainly present as  was illustrated before in
Figs.~\ref{@s+pvol}
and~\ref{@lognorvo}.
For calculations of this type with Gaussian
distributions see Ref.~\cite{Baaquie}.
If the semigroup property was satisfied perfectly, the L\'evy parameter $ \alpha $
would be time independent
as we can see from Eq.~(\ref{Malp}) with
$ \sigma ^2\propto(t_b-t_a) $ and
$ \kappa\propto1/(t_b-t_a) $.
 With the slower falloff  of $ \kappa \propto 1/\sqrt(t_b-t_a)$,
however, $ \alpha $ decreases like $1/\sqrt(t_b-t_a)$.

\begin{figure}[tbhp]
~\\[-11mm]
\unitlength .36mm
\input convsp.tps \\[1cm]
\caption[Left-hand side: Cumulative distributions obtained
from repeated convolution integrals
of the 15-min
distributions.
The data are fitted by \tld{}s.
Right-hand side:
Falloff of kurtosis is slower than the expected  $1/(t_b-t_a)$
from
convolution property]{Left-hand side: Cumulative distributions obtained
from repeated convolution integrals
of the 15-min distribution
 (from Ref.~\cite{BP}).
  Apart from the far ends of
the tails, the
semigroup property (\ref{@convprtop})
is reasonably well satisfied. Right-hand side:
Falloff of kurtosis is slower than $1/(t_b-t_a)$
 expected
from
convolution property . }
\label{@convsp}\end{figure}

  \subsection{Fokker-Planck-Type Equation}
\label{@FPLEQ}
The $ \delta $-functional
may be represented
by a Fourier integral leading to
\begin{eqnarray}
P(x_b t_b|x_at_a)=
\int {{\cal D}  \eta }
\int \frac{{\cal D}  p }{2\pi}
\exp\left\{ \int _{t_a}^{t_b}dt\left[i p(t)\dot x(t)-ip(t) \eta (t)- \tilde H(  \eta  (t))\right] \right\}
\label{@18spiLP1}       .
\end{eqnarray}
Integrating out the noise variable
$ \eta (t)$
amounts to performing the inverse Fourier transform
(\ref{@deftrld}) at each instant of time
and we obtain
\begin{eqnarray}
P(x_b t_b|x_at_a)=
\int \frac{{\cal D}  p }{2\pi}
\exp\left\{ \int _{t_a}^{t_b}dt\left[i p(t)\dot x(t)- H(  p  (t))\right] \right\}
\label{@18spiLP2}      .
\end{eqnarray}
Integrating over all $x(t)$ with  fixed end points
enforces a constant momentum
along the path, and we remain with a single
integral
\begin{eqnarray}
P(x_b t_b|x_at_a)=
\int \frac{{d}  p }{2\pi}
\exp\left[ -({t_b}-{t_a}) H(  p)+ip(x_b-x_a) \right]
\label{@18spiLP3}      .
\end{eqnarray}
The Fourier integral can now be performed and we obtain, for a \tld:
\begin{eqnarray}
P(x_b t_b|x_at_a)=\tilde L_{ \sigma ^2(t_b-t_a)}^{(\mu, \alpha) )}(x_b-x_a)
\label{@18spiLP4}  .
\end{eqnarray}
The result is therefore a \tld{}
of increasing width.
All expansion coefficients $c_n$ of $H(p)$
in Eq.~(\ref{@cumuatsH}) receive the same
factor $t_b-t_a$, which has the consequence that
the  kurtosis $ \kappa =c_4/c_2^2$
decreases
inversely proportional to
$t_b-t_a$. The distribution becomes increasingly Gaussian
with increasing time, as a manifestation of the central limiting theorem
\iems{central+limiting+theorem}
of statistical mechanics.
This is in contrast to the pure L\'evy distribution
which has no finite width and therefore
maintains its power falloff
at large distances.

From the Fourier representation
(\ref{@18spiLP3}) it is easy to prove that this probability
satisfies a Fokker-Planck-type equation
\begin{eqnarray}
\partial _tP(x_b t_b|x_at_a)=-H(-i\partial _x)
P(x_b t_b|x_at_a).
\label{@FPDEL}\end{eqnarray}
Indeed, the general solution $\psi(x,t)$
of this differential equation
with the initial condition $\psi(x,0)$
is given by the path integral generalizing (\ref{@18spiLP})
\begin{eqnarray}
\psi(x,t)=
\int {{\cal D}  \eta }
\exp\left[ -\int _{t_a}^{t_b}dt \, \tilde H( \eta (t))\right]
 \psi\left( x-\int_{t_a}^t dt' \eta(t')\right).
\label{@18spiLPg}
\end{eqnarray}
To verify  that this
satisfies indeed
the Fokker-Planck-type equation  (\ref{@FPDEL})
we consider $\psi(x,t)$ at a slightly later time
$t+ \epsilon $
and expand
\begin{eqnarray}~~~\!\!\!\!\!\!\!\!\psi(x,t+ \epsilon )~~~
&&\!\!\!\!\!\!\!\!\!\!\!\!\!\!\!\!\!\!\!\!\!\!\!\!\!\!\!\!\!\!~~~~~
\hspace{9mm}=
\int {{\cal D}  \eta }
\exp\left[ -\int _{t_a}^{t_b}dt \, \tilde H( \eta (t))\right]
 \psi
\left( x-\int_{t_a}^t dt' \eta(t')-\int_{t}^{t+ \epsilon } dt' \eta(t')\right).
~~~~\nonumber \\
&&\!\!\!\!\!\!\!\!\!\!\!\!\!\!\!\!\!\!\!\!\!\!\!\!\!\!\!\!\!\!~~~~~
\hspace{9mm}
=
\int {{\cal D}  \eta }
\exp\left[ -\int _{t_a}^{t_b}dt \, \tilde H( \eta (t))\right]  \left\{
 \psi
\left( x-\int_{t_a}^t dt' \eta(t')\right)
\right.\nonumber \\
&&\!\!\!\!\!\!\!\!\!\!\!\!\!\!\!\!\!\!\!\!\!\!\!\!\!\!\!\!\!\!~
\hspace{9mm}
~~~~
- \left.~~\psi '
\left( x-\int_{t_a}^t dt' \eta(t')\right)
\int_{t}^{t+ \epsilon } dt' \eta(t')
\right.\nonumber \\
&&\!\!\!\!\!\!\!\!\!\!\!\!\!\!\!\!\!\!\!\!\!\!\!\!\!\!\!\!\!\!~
\hspace{9mm}
~~~~+\left.\frac{1}{2} \psi ''
\left( x-\int_{t_a}^t dt' \eta(t')\right)
\int_{t}^{t+ \epsilon } dt_1 dt_2\,\eta(t_1)
\eta(t_2)
\label{@18spiLPgl1}
\right. \\
&&\!\!\!\!\!\!\!\!\!\!\!\!\!\!\!\!\!\!\!\!\!\!\!\!\!\!\!\!\!\!~
\hspace{9mm}
~~~~-\left.\frac{1}{3!} \psi '''
\left( x-\int_{t_a}^t dt' \eta(t')\right)
\int_{t}^{t+ \epsilon } dt_1 dt_2dt_3\,\eta(t_1)
 \eta(t_2)
 \eta(t_3)
\right.\nonumber \\
&&\!\!\!\!\!\!\!\!\!\!\!\!\!\!\!\!\!\!\!\!\!\!\!\!\!\!\!\!\!\!~
\hspace{9mm}
~~~~+\left.\frac{1}{4!} \psi ^{(4)}
\left( x-\int_{t_a}^t dt' \eta(t')\right)
\int_{t}^{t+ \epsilon } dt_1 dt_2dt_3dt_4\,\eta(t_1)
 \eta(t_2)
 \eta(t_3)
 \eta(t_4)   +\dots~
\right\}.\nonumber
\end{eqnarray}
Inserting here
correlation functions
(\ref{@correlfu1})--(\ref{@correlfu})
we obtain
\begin{eqnarray}
\psi(x,t+ \epsilon )
&=&
\int {{\cal D}  \eta }
\exp\left[ -\int _{t_a}^{t_b}dt \, \tilde H( \eta (t))\right]
\nonumber \\
&\!\!\!\!\times&\left[
-\epsilon c_1\partial _x
+\left(\epsilon c_2 + \epsilon ^2c_1\right)\frac{1}{2} \partial^2 _x
- \left(\epsilon c_3+ 3\epsilon ^2c_2c_1\right)\frac{1}{3!}\partial _x^3
\right.
\label{@18spiLPgl1w}
 \\
&\!\!\!\!+&\left.
( \epsilon c_4+ \epsilon ^24c_3c_1+ \epsilon ^23c_2^2
+ \epsilon ^3c_2c_1^2+ \epsilon ^4c_1^2)
\frac{1}{4!}\partial ^4_x
+\dots\right]
 \psi
\left( x-\int_{t_a}^t dt' \eta(t')\right)
.   \nonumber
\end{eqnarray}
In the limit $ \epsilon \rightarrow 0$, only the linear terms
in $ \epsilon $ contribute,
which are all due
to the connected
parts of the correlation functions of $ \eta (t)$.
The differential operators in the brackets
can now be pulled out of the integral and we find
 the differential equation
\begin{eqnarray}
\!\!\!\!\!\!\!\!~\partial _t\psi(x,t )
&=&
\left[- c_1\partial _x
+ c_2 \frac{1}{2} \partial^2 _x
- c_3 \frac{1}{3!} \partial^3 _x
+c_4
\frac{1}{4!}\partial ^4_x+\dots\right]
 \psi
\left( x,t\right)
.
\label{@18spiLPgl1w3}
\end{eqnarray}
 We now
 replace
$c_1\rightarrow r_{x}$
 and
express
using
 (\ref{@cumuatsHar})
 the
differential operators in  brackets
as Hamiltonian operator
$-H_{r_{x}}(-i\partial _x)$. This leads to
 the Schr\"odinger-like equation
\begin{eqnarray}
~~~~~\partial _t\psi(x,t )
&=&
-H_{r_{x}}(-i\partial _x)
\,\psi\left( x,t\right)
.
\label{@18spiLPgl1w3H}
\end{eqnarray}

Note that due to the many derivatives in $H(i\partial _x)$, this equation is
in general non-local.

By a similar procedure  as in the derivation of Eq.~(\ref{@18spiLPgl1w3})
it is possible
to derive a generalization
of Ito's rule (\ref{@Itosrule})
to functions of noise variable with non-Gaussian
distributions.
Thus
we expand
$f( x(t+ \epsilon ))$ as in (\ref{@Itonew200})
where $x(t)$
satisfies now
 the stochastic differential equation
 $\dot x(t)= \eta (t)$
with a nonzero expectation value $\langle  \eta (t)\rangle =c_1$.
In contrast to the previous
evaluation of
(\ref{@Itonew200})         with Gaussian noise,
which had to be carried out only up to second order
in the noise variable,
we must now keep  {\em all\/} orders.
Evaluating the noise averages of the multiple integrals on the right-hand side
using the correlation functions
(\ref{@correlfu1})--(\ref{@correlfu}),
we find
 the
time derivative of the expectation value of an arbitrary function
of the fluctuating variable
$x(t)$
\begin{eqnarray}\!\!\!\!\!\!\!\!\!\!\!\!\!\!\!\! \!\!
 \!\langle f( x(t\!+\! \epsilon )) \rangle
\! &=&
\langle  f(x(t))\rangle +
\langle f'(x(t)) \rangle \epsilon c_1
+\frac{1}{2}\langle f''(x(t))\rangle ( \epsilon c_2\!+\! \epsilon ^2c_1^2)\nonumber \\
\! &&
~~~~\,~~~~~~~~\!+\frac{1}{3!}\langle f^{(3)}(x(t))\rangle
( \epsilon c_3
+ \epsilon ^2c_2c_1+
 \epsilon ^3c_1^3)
+\dots  \nonumber \\
&\!\!\!\hspace{1pt}=\!\!\!& \epsilon \left[- c_1\partial _x
+ c_2 \frac{1}{2} \partial^2 _x
- c_3 \frac{1}{3!} \partial^3 _x
+\dots\right]\langle f(x(t))\rangle+{\cal O}( \epsilon ^2).
\label{@Itonew0x}\end{eqnarray}
After the replacement
$c_1\rightarrow r_{x}$
the function
$f(x(t))$ obeys
therefore
 the following equation:
\begin{eqnarray}
 \langle \dot f( x(t))\rangle =  -H_{r_{x}}(i\partial _x)\langle f(x(t))\rangle .
\label{@Itonewex}\label{@Itonew}\end{eqnarray}
Taking out the lowest-derivative term this takes a form
\begin{eqnarray}
 \langle \dot f( x(t))\rangle = \langle  \partial _xf(x(t))\rangle
\langle \dot x(t)\rangle
-\bar H_{r_{x}}(i\partial _x)\langle f(x(t))\rangle .
\label{@Itonewbar}\end{eqnarray}
In  postpoint time slicing, this may be viewed as the expectation value of
the stochastic differential equation
\begin{eqnarray}
 \dot f( x(t)) =  \partial _xf(x(t))
\dot x(t)
-\bar H_{r_{x}}(i\partial _x)f(x(t)) .
\label{@Itonewex}\label{@Itonewb}\end{eqnarray}
This is the direct generalization of
 \ind{Ito's rule}
 (\ref{@Itosrule}).
\comment{and this may
be thought of a the expectation value of the
 stochastic equation
\begin{eqnarray}
 \dot f( x) = f'(x)\dot x -\bar H(i\partial _x)f(x),
\label{@Itonew}\end{eqnarray}
generalizing
Ito's
rule (\ref{@Itosrule}). }

For an exponential  function
$f(x)=e^{Px}$,
this becomes
%
\begin{eqnarray}
 \frac{d}{dt}  e^{Px(t)} =\left[ \dot x(t) -H_{r_{x}}(i P)\right]e^{Px(t)}
 .
\label{@}\end{eqnarray}
\comment{a result closely related to  (\ref{@modwick}).}

\comment{Going back to the original
stochastic differential equation
 (\ref{@stdeqet})
in which
the average linear growth $r_{x}$
is shown explicitly
as
\begin{equation}
\dot x(t)=r_{x}+  \eta (t),
\label{@ratexvn}\end{equation}
with a vanishing noise average,
the relation (\ref{@Itonewex})
becomes
\begin{eqnarray}
\langle  \dot f( x)\rangle =
r_{x}\langle f'(x)\rangle
-\bar H(i\partial _x)\langle f(x)\rangle .
\label{@Itonewn}\end{eqnarray}
}
As a consequence of this equation for $P=1$,
the rate $r_S$ with which a
stock price
$S(t)=e^{x(t)}$
grows on the average according to formula
(\ref{@stochdequsto}) is
now related to
$r_{x}$   by
\begin{equation}
r_S=
r_{x}-\bar H(i)=
r_{x}-[H(i)-iH'(0)]=
-H_{r_{x}}(i),
\label{@ratexvnru}\end{equation}
which replaces
  the simple
Ito relation
$r_{S}=r_{x}+\sigma ^2/2$ in Eq.~(\ref{@itoreln}).
Recall
the definition
$\bar H(p)\equiv H(p)-H'(0)p$
in Eq.~(\ref{@cumuatsHar}).
The corresponding  generalization of  the left-hand part of
Eq.~(\ref{@stochdequstox})
 reads
\begin{equation}
\frac{\dot S}{S}
=\dot x(t)
-\bar H(i)
=
\dot x(t)
-[H(i)-iH'(0)]
=\dot x(t) -r_{x}
-H_{r_{x}}(i)
.
\label{@stochdequstoxn}\end{equation}
The forward price of a stock must therefore be calculated with
the generalization of formula
(\ref{@forwS}):
\begin{equation}\!\!
\langle S(t)\rangle =S(0)e^{r_St}=
S(0)\langle e^{r_{x}t+\int_0^t dt'\, \eta (t')}\rangle
=S(0)e^{-H_{r_{x}}(i)t}
=S(0)e^{\{r_{x}-[H(i)-iH'(0)]t\}}.
\label{@forwSn}\end{equation}

Note that
e may derive
the differential equation of an arbitrary function $f(x(t))$
in Eq.~(\ref{@Itonew}) from a simple
mnemonic rule, expanding sloppily
\begin{eqnarray}  \!\!\!\!\!\!\!\!\!\!\!\!\!\!\!\!\!\!\!\!
f(x(t+ dt ))&=&
f(x(t)+\dot x dt)=
f(x(t))+f'(x(t))\dot x(t) dt+
\frac{1}{2}f''(x(t))\dot x^2(t)dt^2
\nonumber \\&\!\!\!+&
\frac{1}{3!}f^{(3)}(x(t))\dot x^3(t)dt^3
+\dots~,
\label{@20.364n}\end{eqnarray}
and replacing
\begin{eqnarray}  \!\!\!\!\!\!\!\!\!\!
\langle\dot x(t)\rangle dt
\rightarrow c_1 dt,~~ ~
\langle\dot x^2(t)\rangle dt^2
\rightarrow c_2 dt,~~  ~
\langle\dot x^3(t)\rangle dt^2
\rightarrow c_3 dt,\dots\,.~~~~
\label{@Itosruleg}\end{eqnarray}

\section[]{Conclusion}

The new stochastic calculus
developed in this paper
should be useful for
estimating financial risks
of a variety of investments.
In particular,
it will help in
developing simple methods
to estimating
fair option prices
for stocks with non-Gaussian fluctuations.
More details can be found in
the textbook Ref.~\cite{PI}.

~\\           ~\\
~\\           ~\\
Acknowledgment

The author is grateful to Marc Potters for discussions.
He also  thanks Axel Pelster,  Flavio Nogueira,
Ernst Eberlein, Jan Kallsen, Martin Schweizer, Peter Bank, and
E.C. Chang
for useful comments.

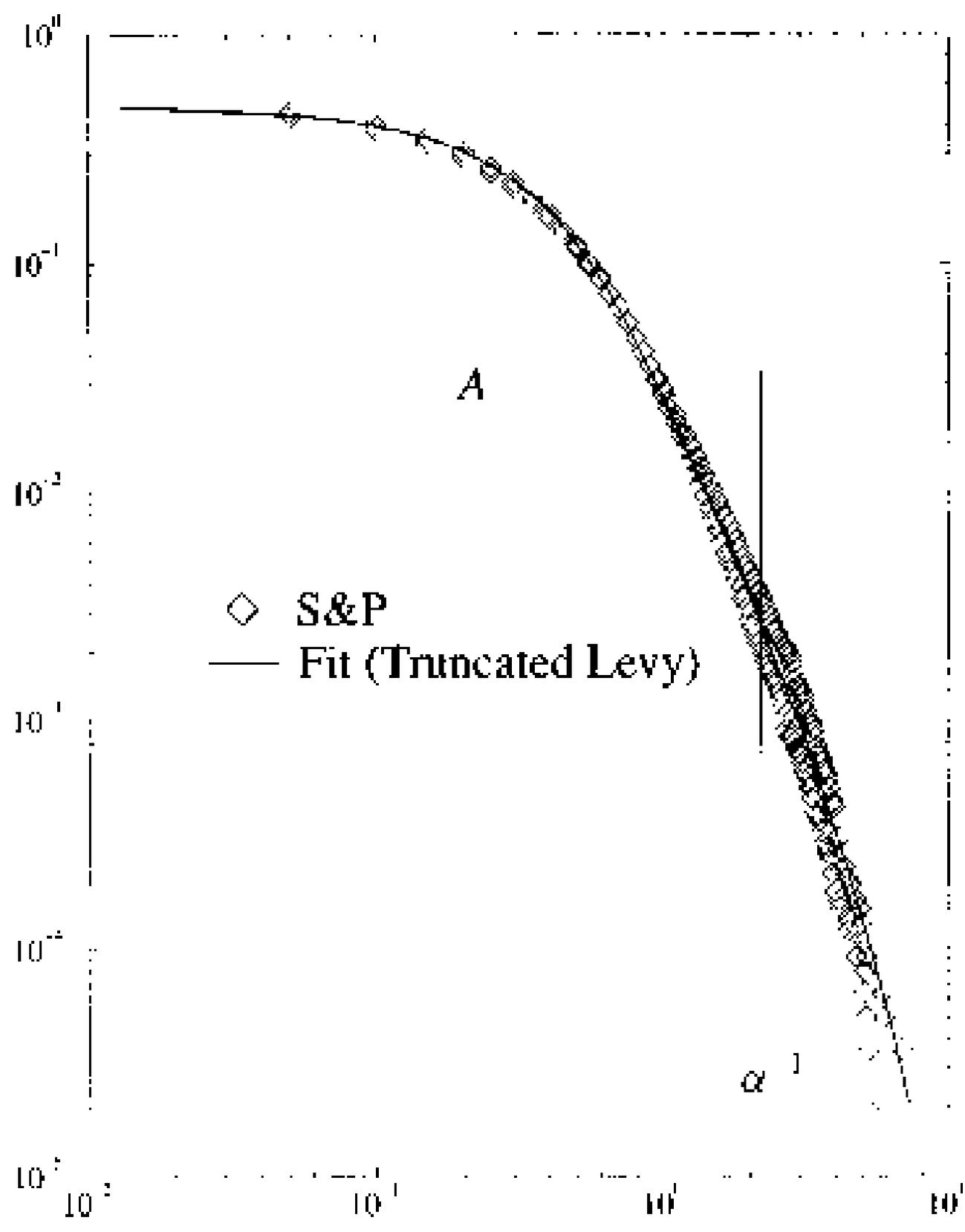
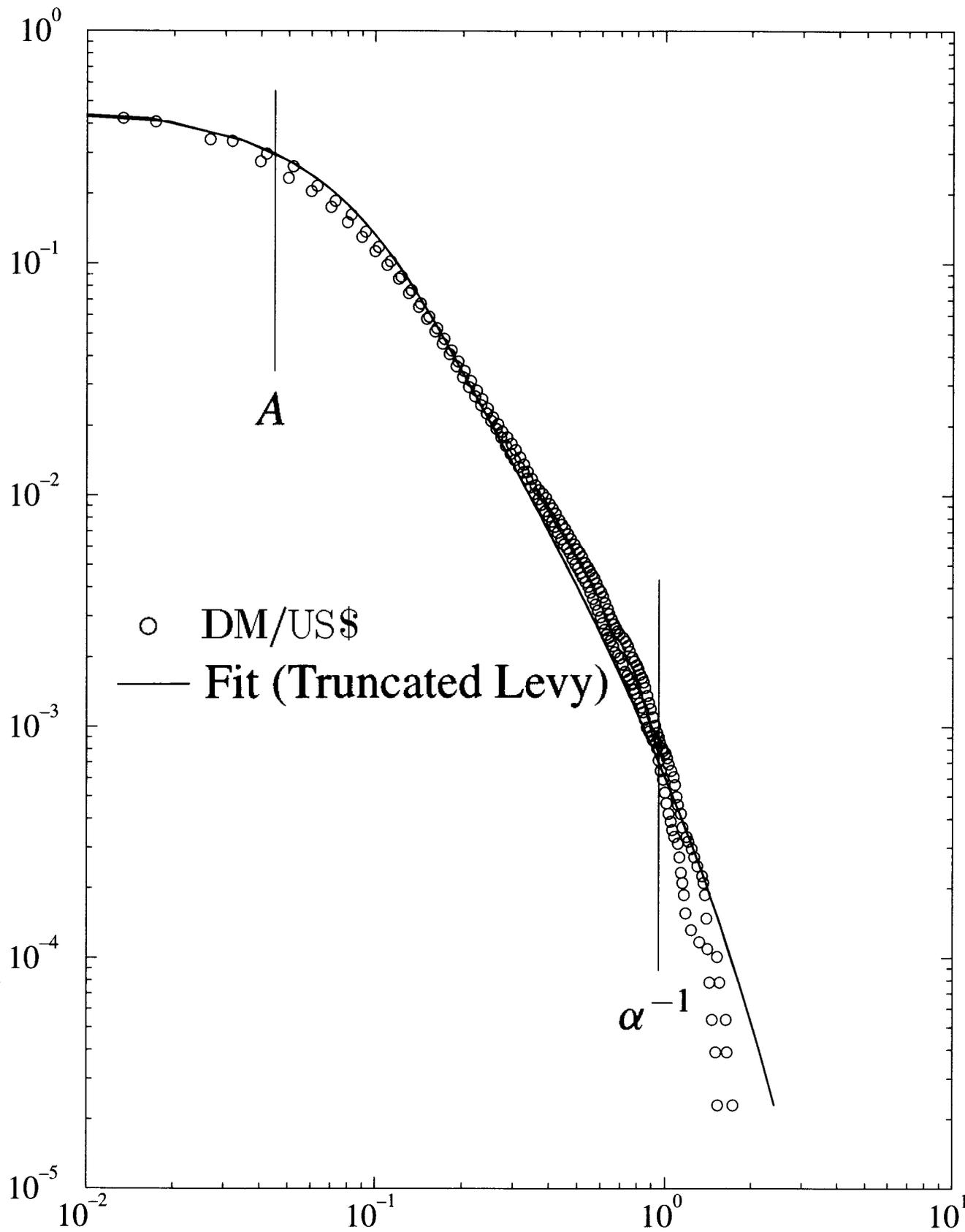
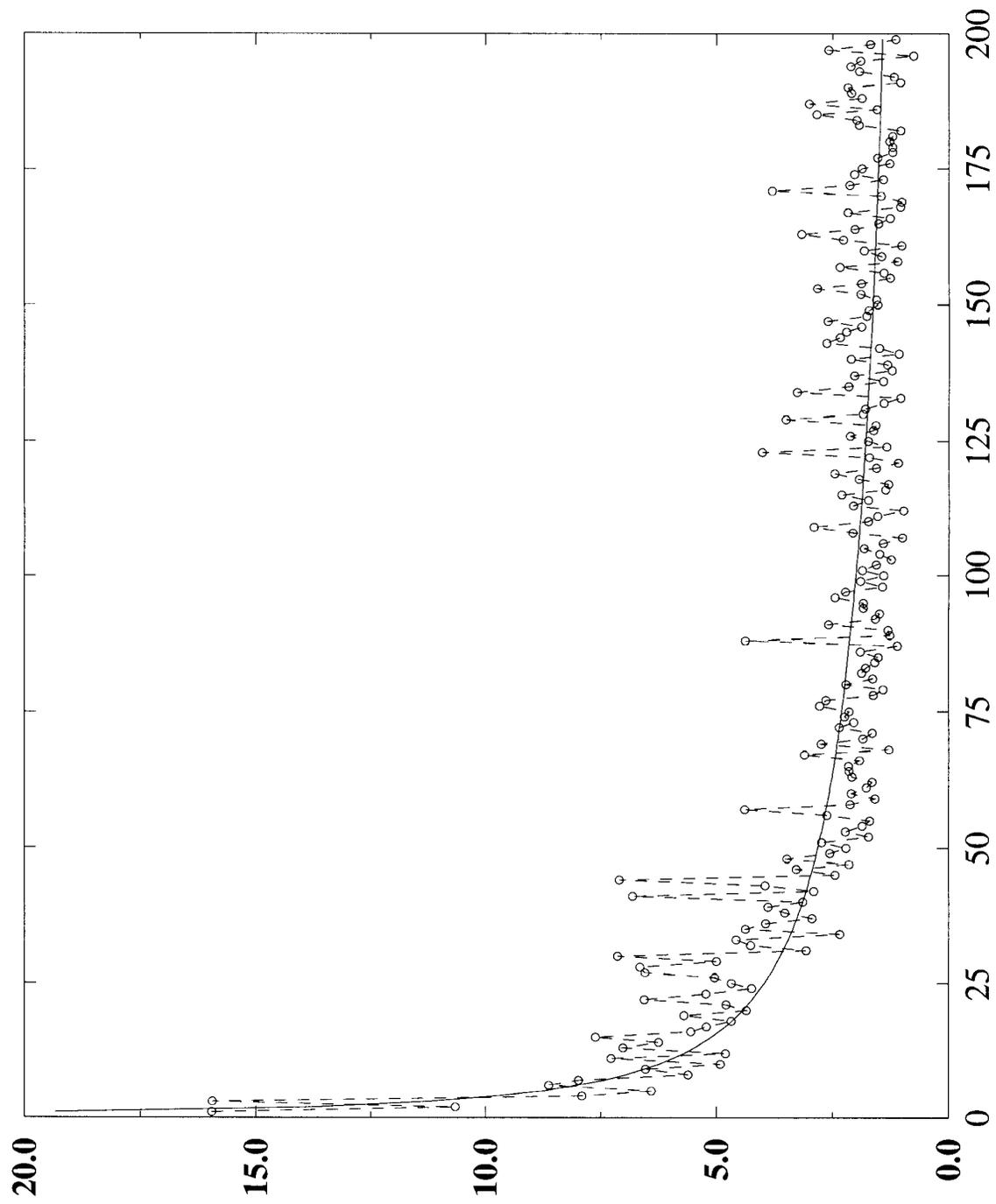
\end{document}